\documentclass[journal=jctcce,manuscript=article]{achemso}

\usepackage{chemformula} 
\usepackage[T1]{fontenc} 
\usepackage{titlesec}
\usepackage{amsmath}
\usepackage{amssymb}
\usepackage{mathtools}
\usepackage{multirow}
\usepackage{float}
\usepackage{graphicx}

\DeclarePairedDelimiter\ket{\lvert}{\rangle}

\author{Ruiyi Zhou$^\dagger$}
\author{Qiujiang Liang$^\dagger$}
\author{Jun Yang}
\email{juny@hku.hk}
\affiliation[The University of Hong Kong]
{Department of Chemistry, The University of Hong Kong, 
Hong Kong SAR, P.R. China}

\title[]
  {
A complete OSV-MP2 analytical gradient theory for molecular structure and dynamics simulations
\footnote{{Equal contributions}}
   }

\graphicspath{{./figures/}}
\begin{document}

%
%

\begin{abstract}

We propose an exact algorithm for computing the analytical gradient within the
framework of the orbital-specific-virtual (OSV) second-order M{\o}ller-Plesset
(MP2) theory in resolution-of-identity (RI) approximation.  We implement the
relaxation of perturbed OSVs through the explicit constraints of the
perturbed orthonormality, the perturbed diagonality and the perturbed eigenvalue
condition. We show that the rotation of OSVs within the retained OSV
subspace makes no contribution to gradients, as long as the unperturbed
Hylleraas energy functional reaches minimum. The OSV relaxation is solved as the
perturbed non-degenerate eigenvalue problem between the retained and discarded
OSV subspaces.  The detailed derivation and preliminary implementations for
gradient working equations are discussed. The coupled-perturbed localization
method is implemented for meta-L\"owdin localization function.  The numerical
accuracy of computed OSV-MP2 gradients is demonstrated for the geometries of
selected molecules that are often discussed in other theories.  From OSV-MP2
with the normal OSV selection, the canonical RI-MP2/def2-TZVP gradients can be
reproduced within $10^{-4}$ a.u.  The OSV-MP2/def2-TZVPP covalent bond lengths,
angles and dihedral angles are in good agreement with canonical RI-MP2
structures by 0.017 pm, $0.03^\circ$ and $0.2^\circ$, respectively.  No
particular accuracy gains have been observed for molecular geometries compared
to the recent local pair-natural-orbital MP2 by using the predefined orbital
domains.  Moreover, the OSV-MP2 analytical gradients can generate atomic forces
that are utilized to drive the Born-Oppenheimer molecular dynamics (BOMD)
simulation for studying structural and vibrational properties with respect to
OSV selections.  By performing the OSV-MP2 $NVE$ BOMD calculation using the
normal OSV selection, the structural and vibrational details of protonated water
cations are well reproduced.  The 200 picoseconds $NVT$ well-tempered
metadynamics at 300 K has been simulated to compute the OSV-MP2 rotational free
energy surface  of coupled hydroxyl and methyl rotors for ethanol molecule.
\end{abstract}
\section{INTRODUCTION}

\textit{Ab-initio} electronic structure theory has been significantly
progressed with many theoretical and algorithmic developments.  Reduced-scaling
post-Hartree-Fock methods are now capable of efficiently computing molecular
systems of substantially increased size by managing trade-offs between the
accuracy that can be achieved and the resource that can be
accessed\cite{zalesny2011linear, gordon2017fragmentation}.  The reduced-scaling
techniques are often based on the unique strength of the spatial locality,
i.e., the short-range behaviour of electron correlation that emerges as the
size of a system increases.  Many schemes have been devised and implemented to
compute the energies of large molecules by operating on a sufficiently accurate
and reduced subset of of Hilbert space  in which an approximate wavefunction
can be efficiently represented, manipulated and
stored\cite{martinez1994pseudospectral,ayala1999linear,reynolds1996local,lee2000closely}.

The locality of dynamic electron correlation was introduced by
Pulay\cite{pulay1983localizability} and initially implemented by
S{\ae}b\o\cite{saebo1985local,saebo1987fourth,saebo1993local}.  This has led to
a fruitful variety of wavefunction representations by which the unphysical
steep computational scaling can be drastically diminished.  Notably, a
hierarchy of M{\o}ller-Plesset perturbation and coupled-cluster (CC) methods
has been developed by employing projected atomic orbitals (PAO) by Werner,
Sch{\"u}tz and coworkers
\cite{hampel1996local,schutz2000local,schutz2001low,schutz2002low,schutz2002new,werner2011efficient},
pair-nature-orbitals (PNOs) pioneered by Meyer et al.
\cite{meyer1971ionization,ahlrichs1975pno} and revitalized by
Neese\cite{neese2009efficient}, and orbital-specific-virtuals (OSVs) by Chan
\cite{yang2011tensor,kurashige2012optimization,yang2012orbital,schutz2013orbital}.
By construction, the PNO and OSV are both inherently local and specific to a
single  orbital pair and orbital, respectively.  The hybrid near-linear-scaling
PNO-MP2 and PNO-CCSD schemes by mixing PAO/OSV/PNO have been demonstrated to
further reduce the number of PNOs that are required to compress the cluster
operators by employing  PAO or OSV as an intermediate stage
\cite{krause2012comparison,riplinger2013efficient,riplinger2013natural,schmitz2013scaling,
werner2015scalable}.  The hybrid PNO schemes ensure the most compact virtual
space for recovering a certain percentage of correlation energy.  In addition,
explicitly correlated CCSD(T) methods in the PNO framework have been developed
to reduce basis set error
\cite{schmitz2014explicitly,schmitz2016perturbative,pavovsevic2016sparsemaps,pavovsevic2017sparsemaps,ma2017scalable,ma2018scalable}.
Open-shell PNO-CCSD\cite{saitow2017new}, PNO variants of state-specific
multi-reference perturbation and CC theories
\cite{demel2015local,guo2016sparsemaps,menezes2016local,brabec2018domain,lang2019perturbative},
as well as PNO-based
EOM-CC2\cite{helmich2013pair}/CCSD\cite{frank2018pair,peng2018state},
CIS(D)\cite{helmich2011local}, ADC(2)-x\cite{helmich2014pair} for excited
states in both state-specific and state-average approaches have also been
implemented and demonstrated.

A wide range of chemistry problems, such as molecular geometries, reaction
pathways, thermal and spectroscopic properties, and so on, involves the
physical motion of atoms.  In essence, these molecular properties require an
efficient computation of analytical energy
gradients\cite{pulay1969ab,yamaguchi1994new} with respect to relaxations of
molecular orbitals and/or other parameters of a deterministic electronic
wavefunction.  Apparently, analytical gradient techniques are highly specific
to the way in which wavefunction of the system is constructed.  In the past
decades, for instance, for computing analytical gradients with manageable
cost-accuracy balance, the implementations have adopted very different
reduced-scaling strategies for the variants of MP2
\cite{weigend1997ri,lee2000closely,hattig2006distributed,lochan2007quartic,
distasio2007analytical,distasio2007improved,schweizer2008atomic,kristensen2012molecular}
and CC methods
\cite{adamowicz1984analytical,fitzgerald1985analytical,scheiner1987analytic,salter1989analytic,
scuseria1991analytic,hald2003lagrangian,bozkaya2017analytic}.

For applications to large molecules, the PAO-based analytical gradients have
been established for the local MP2\cite{el1998analytical,schutz2004analytical},
CC2\cite{ledermuller2014local} and CCSD\cite{rauhut2001analytical} models.  In
the context of the more recent PNO and OSV schemes, the implementation of PNO-
or OSV-based analytical gradients is more limited, primarily due to the
complexity of PNO or OSV related approximations.  The performance of the
simulated PAO-, PNO- and OSV-based CCSD  for non-resonant optical properties
was assessed by Crawford and coworker using linear response
theory\cite{mcalexander2015comparison}.  The PNO-MP2 and PNO-CCSD analytical
energy gradients were developed by H\"attig\cite{frank2017pno} and
Neese\cite{datta2016analytic}, respectively, without accounting for the
relaxation of PNOs.  Most recently, the PNO relaxation  problem was
circumvented for the domain-based local PNO-MP2 (DLPNO-MP2) by enforcing a
block-diagonal semi-canonical external pair density matrix which assumes zero
off-blocks between the retained and discarded PNO
orbitals\cite{pinski2018communication,pinski2019analytical}, making the PNO-MP2
energy invariant to the rotation among the kept PNO orbitals.

In the present work, we turn our attention to developing the exact OSV-MP2
analytical gradient theory for its much simpler way of constructing OSVs
relative to the hybrid PNOs in DLPNO-MP2 model.  
Here, we implement the OSV relaxation explicitly as a perturbed  eigenvalue
problem by using both orthonormality and eigenvalue conditions for perturbed
OSVs. We also show that the degenerate eigenvalue issue that may break OSV
relaxation\cite{pinski2018communication,pinski2019analytical} does not occur for
reasons that will be described in our formalism and implementation. 
The resulting OSV relaxation vectors may be further scrutinized in a way 
that their intrinsic sparsity can be explored to pre-select
important OSV relaxations making most contributions to OSV-MP2 gradients. 
In addition,
OSV-MP2 was shown to produce smooth potential energy curve with respect to
molecular structures even with small pair domains\cite{yang2011tensor}.
This is essential to efficient simulations of Born-Oppenheimer molecular dynamics
(BOMD)\cite{russ2004potential}. 


The paper is organized as follows. In Sec. II we describe the details of our
OSV-MP2 analytical gradient theory and implementation.  We implemented the
algorithm in a standalone Python program, and \texttt{PYSCF}\cite{sun2018pyscf}
has been used for obtaining the one- and two-electron integrals,  their
derivatives and the RHF reference wavefunction.  In Sec. III we compute the
optimized molecular structures and assess the accuracy of the  OSV-MP2
analytical gradients with respect to the  selection parameters for OSVs and
orbital pairs.  The results are also compared with  canonical  MP2 and
DLPNO-MP2 results available in the literature.  In Sec. IV, we carry out the
\textit{ab-initio} BOMD simulations
driven by OSV-MP2 analytical gradients.  Illustrative applications of OSV-MP2
metadynamics are demonstrated to protonated water cations and ethanol molecule.
Here our focus is to find out whether the errors due to the OSV approximations
for cost-accuracy trade-off have any significance in reproducing the energy,
structural and vibrational details by BOMD simulations at a finite-temperature.
Our work is concluded in Sec. V.

\section{THEORY AND IMPLEMENTATION}

\subsection{OSV-MP2 wavefunction}

In the previous work by one of the authors,
the OSV-based single-reference local MP2, CCSD and CCSD(T) methods\cite{yang2011tensor,yang2012orbital,schutz2013orbital} 
were developed.
In this section, we briefly review the algorithm for introducing the notations
relevant to the OSV-MP2 gradient algorithm.
We use $i,j,k,\cdots$ to denote the occupied localized molecular orbitals (LMOs),
$a,b,c,\cdots$ canonical virtual MOs,
$\bar\mu_k,\bar\nu_k,\bar\xi_k,\cdots$ the OSVs associated with an occupied MO $k$,
while  $p,q,r,\cdots$ and $\alpha,\beta,\cdots$ pertain to  generic
indices of MOs and atomic orbitals (AOs), respectively.
Here LMOs refer to the spatial orbital basis.
The bra-ket symbol $\langle \cdots \rangle$ is used to evaluate the matrix trace through the discussion.

In OSV ans\"atz, a sparse structure of the amplitudes $\mathbf{T}_{(ij,ij)}$ and the first-order
wavefunction $\ket{\mathbf\Phi_{(ij,ij)}}$ 
can be explored by  constructing  a compact virtual space 
in a transformative OSV adaption to the occupied space 
by associating a set of OSVs $\{\bar\mu_k\}$ with each occupied orbital $k$,
\begin{equation}
  \ket{\bar\mu_k} = \sum_a Q_{a\bar\mu}^{k} \ket a \label{eq:osv}
\end{equation}
The compactness of the OSV space is determined by the tensorial character
of the transformation matrix $\mathbf{Q}_{k}$ for each occupied orbital.
An excellent yet simple choice\cite{yang2011tensor} of $\mathbf{Q}_{k}$ is to 
require its column vector to be the orthonormal eigenvector 
of the MP2 diagonal pair amplitudes $\mathbf{T}_{kk}$ for each  $k$ 
by performing the diagonalization,
\begin{equation}
 \left[ \mathbf{Q}_k^\dagger \mathbf{T}_{kk} \mathbf{Q}_{k} \right]_{\bar\mu_k\bar\nu_k}
  = \omega_{\bar\mu_k} \delta_{\bar\mu\bar\nu}, \label{eq:svd}
\end{equation}
with the orthonormality $\mathbf{Q}_k^\dagger \mathbf{Q}_{k} = \mathbf{1}$.
According to the magnitude of eigenvalues $\omega_{\bar\mu_k}$,
a single parameter $l_{osv}$ is utilized as a measure to select
a set of OSVs pertaining to each occupied orbital $k$ 
by which $\mathbf{T}_{(ij,ij)}$ is solved efficiently
without losing too much accuracy.
The elements of $\mathbf{T}_{kk}$ in Eq.~(\ref{eq:svd}) 
are computed as
\begin{equation}
 [\mathbf{T}_{kk}]_{ab} = \frac{(k a\rvert k b)}{f_{aa}+f_{bb}-2f_{kk}}.
 \label{eq:tkk}
\end{equation}
$f_{kk}, f_{aa}, f_{bb}$ are the diagonal elements of the Fock matrix. 

The OSV wavefunction $\ket{\mathbf{\Phi}_{(ij,ij)}}$ and 
amplitudes $\mathbf{T}_{(ij,ij)}$ are associated with a collated excitation manifold 
in which the occupied orbital $i$ excites to its own OSV set 
$\{\bar\mu_i\}$ ($i\rightarrow\bar\mu_i$) as well as the exchange set 
$\{\bar\nu_j\}$ ($i\rightarrow\bar\nu_j$),
\begin{equation}
 \ket{\mathbf{\Phi}_{(ij,ij)}} = \left( {\begin{array}{cc} \ket{\Phi_{ij}^{\bar\mu_i\bar\nu_i}} &
 \ket{\Phi_{ij}^{\bar\mu_i\bar\xi_j}}\\ \ket{\Phi_{ij}^{\bar\sigma_j\bar\nu_i}} & 
 \ket{\Phi_{ij}^{\bar\sigma_j\bar\xi_j}} \end{array}} \right)^\mathrm{T},\hfill 
  \mathbf{T}_{(ij,ij)} = \left( {\begin{array}{cc} t_{ij}^{\bar\mu_i\bar\nu_i} &
 t_{ij}^{\bar\mu_i\bar\xi_j}\\ t_{ij}^{\bar\sigma_j\bar\nu_i} & t_{ij}^{\bar\sigma_j\bar\xi_j} \end{array}} \right).\label{eq:t2}
\end{equation}
The doubly excited configuration $\ket{\Phi_{ij}^{\bar\mu_k\bar\nu_l}}$ 
is built  through the spin-free excitations operator 
$\hat{E}_i^{\bar\mu_k} = \sum_\sigma \hat{a}^\dagger_{\bar\mu_k\sigma} \hat{a}_{i\sigma}$
in terms of the creation and annihilation operators for all spins 
$\sigma=\uparrow, \downarrow$ acting on the zero-order wavefunction 
$\ket{\Phi_{ij}^{\bar\mu_k\bar\nu_l}} = \hat{E}_i^{\bar\mu_k}\hat{E}_j^{\bar\nu_l} \ket{\Psi^{(0)}}$.
The OSV amplitudes $\mathbf{T}_{(ij,ij)}$ are computed iteratively
by solving the residual equation 
$\mathbf{R}_{(ij,ij)}$ for an $(i,j)$ pair,
\begin{equation}
  \mathbf{R}_{(ij,ij)} = \mathbf{K}_{(ij,ij)} + \sum_k \mathbf{S}_{(ij,ik)} 
  \mathbf{T}_{(ik,ik)} [\delta_{kj} \mathbf{F}_{(ik,ij)}-f_{kj}\mathbf{S}_{(ik,ij)}]
  + [\delta_{ik}\mathbf{F}_{(ij,kj)}-f_{ik}\mathbf{S}_{(ij,kj)}]\mathbf{T}_{(kj,kj)}\mathbf{S}_{(kj,ij)}.
   \label{eq:rij}
\end{equation}
In the OSV basis, $\mathbf{K}_{(ij,ij)}$, $\mathbf{S}_{(ik,ij)}$ and $\mathbf{F}_{(ik,ij)}$
denote the two-electron integrals,
overlap and Fock matrices for an $(i,j)$ pair, respectively.
$\mathbf{A}_{(ij,kl)}$ is adopted to represent a generic composite matrix 
assembled between $\{\bar\mu_i,\bar\nu_j\}$ and $\{\bar\sigma_k,\bar\xi_l\}$ elements as needed. 
In essence, $\mathbf{A}_{(ij,kl)}$  is a projection of $\mathbf{A}$ 
from the canonical virtual MOs to OSVs basis
\begin{equation}
 \mathbf{A}_{(ij,kl)} = 
 \left( 
  {\begin{array}{c} 
    \mathbf{Q}_i^\dagger\\\mathbf{Q}_j^\dagger 
   \end{array}}
 \right)
  \mathbf{A}
 \left( 
  {\begin{array}{cc} 
    \mathbf{Q}_k & \mathbf{Q}_l
   \end{array}}
 \right). 
\label{eq:aijkl}
\end{equation}
Since $\mathbf{A}$ is hermitian in canonical MO basis, permuting $(ij)$ and $(kl)$ pairs
yields the self-adjoint property of $\mathbf{A}_{(ij,kl)}$,
\begin{equation}
\mathbf{A}_{(ij,kl)}^{\dagger}=\mathbf{A}_{(kl,ij)}.\label{eq:aadjoint}
\end{equation}
In the OSV basis, the MP2 Hylleraas\cite{hylleraas1930ea} correlation energy $E_c$ has the following form of Lagrangian,
\begin{equation}
 E_c =  \sum_{ij}\mathbf{\langle}\mathbf{K}_{(ij,ij)}\overline{\mathbf{T}}_{(ij,ij)}\mathbf{\rangle}
   + \mathbf{\langle}\mathbf{R}_{(ij,ij)}\overline{\mathbf{T}}_{(ij,ij)}\mathbf{\rangle}. \label{eq:ec}
\end{equation}
This energy Lagrangian essentially imposes the vanishing residual condition $\mathbf{R}_{(ij,ij)}=0$ 
with the corresponding multiplier 
$\overline{\mathbf{T}}_{(ij,ij)}=2\mathbf{T}_{(ij,ij)}-\mathbf{T}_{(ij,ij)}^\dagger$.
An elimination of the linear dependency in the OSV-concatenated  pair domain is
essential for solving $\mathbf{R}_{(ij,ij)}=0$, and can be effectively carried
out by preconditioning $\mathbf{R}_{(ij,ij)}$ in a transformation made by
nonredundant vectors\cite{yang2011tensor}.

\subsection{Perturbed OSVs  and relaxation}

\subsubsection{OSV orbital rotation}
The OSVs are defined as the eigenvectors $\mathbf{Q}_k$ 
of the semi-canonical MP2 diagonal pair amplitude associated 
with a specific occupied orbital $k$, 
as given in Eqs. (\ref{eq:osv}) and (\ref{eq:svd}).
Upon a perturbation $\lambda$ acting on the system,
the perturbed OSVs can be expanded  exactly
in a linear combination of the complete unperturbed OSV basis $\mathbf{Q}_k^0$, 
with the unknown combination coefficient matrix $\mathbf{O}_k$ 
that must be specific to the occupied orbital $k$ as well,
\begin{equation}
 \mathbf{Q}_k(\lambda) = \mathbf{Q}_k^0 \mathbf{O}_k(\lambda). \label{eq:defq}
\end{equation}  
The exact OSV relaxation 
$\mathbf{Q}_k^{\lambda}=\frac{\partial \mathbf{Q}_k(\lambda)}{\partial \lambda}$ 
is thus given in terms of the relaxation matrix $\mathbf{O}_k^{\lambda}$
\begin{equation}
 \mathbf{Q}_k^{\lambda} = \mathbf{Q}_k^0 \mathbf{O}_k^{\lambda}.
 \label{eq:osvrlx}
\end{equation}  
Given the perturbation $\lambda$, the perturbed OSV amplitudes 
$\mathbf{T}_{(ij,ij)}(\lambda)$ must fulfill 
the perturbed residual equation $\mathbf{R}_{(ij,ij)}(\lambda)=0$,
analogous to Eq.~(\ref{eq:rij}).
The perturbed quantity  $\mathbf{A}_{ij,kl}(\lambda)$ of Eq.~(\ref{eq:aijkl}) exhibits 
a dependence on the perturbation 
and can be evaluated with reference to the unperturbed $\mathbf{A}_{ij,kl}^0$,
\begin{equation}
 \mathbf{A}_{(ij,kl)}(\lambda)  = 
 \left( 
  {\begin{array}{c} 
    \mathbf{O}_i^\dagger\\\mathbf{O}_j^\dagger 
   \end{array}}
 \right)
  \mathbf{A}_{(ij,kl)}^0
 \left( 
  {\begin{array}{cc} 
    \mathbf{O}_k & \mathbf{O}_l
   \end{array}}
 \right). \label{eq:ptaijkl}
\end{equation}
Using the OSV relaxation matrix in Eq.~(\ref{eq:osvrlx}),
the OSV derivative in $\mathbf{A}_{(ij,kl)}$ is therefore
\begin{equation}
 \mathbf{A}_{(ij,kl)}^{\{\lambda\}} = \mathbf{O}_{ij}^{\dagger\lambda}
  \mathbf{A}_{(ij,kl)}^0 +\mathbf{A}_{(ij,kl)}^0 \mathbf{O}_{kl}^{\lambda}
 \label{eq:aijklgrad}
\end{equation}
with the curly brackets $\{\}$ specifying the derivatives of OSVs
accounting for the OSV relaxation.
Here we introduce an OSV pair-specific relaxation matrix $\mathbf{O}_{kl}^{\lambda}$ 
for $(k,l)$ pair in a block diagonal form,
\begin{equation} \label{eq:okl}
  \mathbf{O}_{kl}^{\lambda} = diag (\mathbf{O}_k^{\lambda}, \mathbf{O}_l^{\lambda})
\end{equation}
The perturbed OSVs for each orbital must be always orthonormal 
\begin{equation}
 \mathbf{Q}_k^\dagger(\lambda)\mathbf{Q}_k(\lambda)=\mathbf{1}  \label{eq:ptorth}
\end{equation}
which implies that the OSV relaxation matrix $\mathbf{O}_k^{\lambda}$ must be antisymmetric,
\begin{equation}
  \mathbf{O}_k^{\dagger\lambda} + \mathbf{O}_k^{\lambda} = \mathbf{0}. 
\end{equation}

\subsubsection{OSV relaxation as perturbed non-degenerate eigenvalue problem}

Assuming real values of the antisymmetric  $\mathbf{O}_k$, 
all diagonal elements of the OSV relaxation matrix must vanish 
\begin{equation}
[\mathbf{O}_k^{\lambda}]_{\bar\mu\bar\mu}=0. \label{eq:qrspdia}
\end{equation}
Now we discuss an approach in which the off-diagonal 
$\mathbf{O}_k^{\lambda}$ can be explicitly solved 
based on the perturbation analysis\cite{trefethen1997numerical,saad2011numerical} 
to the perturbed eigenvalue problem as
\begin{equation}
 \mathbf{T}_{kk}(\lambda) \mathbf{Q}_{k}(\lambda)
  = \mathbf{Q}_{k}(\lambda)\mathbf{\Omega}_k(\lambda) . \label{eq:ptsvd}
\end{equation}
with  $\mathbf{\Omega}_k(\lambda)=diag\left [\omega_1(\lambda),\omega_2(\lambda),\cdots\right ]$ 
the diagonal eigenvalue matrix.
Differentiating the above equation, we arrive at
\begin{equation}
 \mathbf{T}_{kk}^\lambda \mathbf{Q}_{k}^0 + \mathbf{T}_{kk}^0 \mathbf{Q}_{k}^0 \mathbf{O}_k^{\lambda}
  = \mathbf{Q}_{k}^0\mathbf{\Omega}_k^{\lambda} + \mathbf{Q}_{k}^0\mathbf{O}_k^{\lambda}\mathbf{\Omega}_k^0.
\end{equation}
Multiplying $\mathbf{Q}_k^{0\dagger}$ onto both sides and using the OSV orthonormality,
there is
\begin{equation}
 \mathbf{Q}_k^{0\dagger}\mathbf{T}_{kk}^\lambda \mathbf{Q}_{k}^0 + \mathbf{\Omega}_k^0\mathbf{O}_k^{\lambda}
  = \mathbf{\Omega}_k^{\lambda}  + \mathbf{O}_k^{\lambda}\mathbf{\Omega}_k^0. \label{eq:svdrsp}
\end{equation}
The derivative $\mathbf{T}_{kk}^{\lambda}$ gives
the relaxation of semi-canonical MP2 diagonal amplitudes upon a perturbation.
However, since the canonicality 
$f_{ij}(\lambda)=f_{ii}(\lambda)\delta_{ij}$  
and $f_{ab}(\lambda)=f_{aa}(\lambda)\delta_{ab}$  
does not necessarily hold and in fact is  not required in general
for a perturbed Fock matrix,
$\mathbf{T}_{kk}^{\lambda}$ can not be evaluated directly by
taking the derivative of Eq.~(\ref{eq:tkk}).
Instead, it must be computed by differentiating the MP2 residual equation 
assuming the generic Fock matrix  for a diagonal $kk$ pair,
which leads to the following expression 
\begin{equation}
 [\mathbf{T}_{kk}^{\lambda}]_{ab} = 
   \frac{\left[
           \mathbf{K}_{kk}^{\lambda}
          + \mathbf{T}_{kk} \mathbf{F}^{\lambda} 
          + \mathbf{F}^{\lambda}\mathbf{T}_{kk}
          - 2\mathbf{T}_{kk}f_{kk}^{\lambda}\right]_{ab}}
  {f_{aa}+f_{bb}-2f_{kk}}.\label{eq:tkkrsp}
\end{equation}
Above,  canonical
$\mathbf{K}_{kk}^{\lambda}$ and $\mathbf{F}^{\lambda}$
are composed of the derivatives 
with respect to both AOs ($\lambda$) and MOs [$\lambda$] 
of the exchange integral and Fock matrix, respectively, for instances,
\begin{eqnarray}
\mathbf{K}_{kk}^{\lambda} & = & \mathbf{K}_{kk}^{(\lambda)}+\mathbf{K}_{kk}^{[\lambda]} \\
\mathbf{F}^{\lambda} & = & \mathbf{F}^{(\lambda)} + \mathbf{F}^{[\lambda]} \\
f_{kk}^{\lambda} & = & f_{kk}^{(\lambda)}+f_{kk}^{[\lambda]}.
\end{eqnarray}
Here the MO-specific derivatives $\mathbf{K}_{kk}^{[\lambda]}$ and $\mathbf{F}^{[\lambda]}$ 
are given later according to Eqs.~(\ref{eq:fgradmo}).

The diagonal part of Eq.~(\ref{eq:svdrsp}) yields the relaxation of eigenvalues,
\begin{equation}
 \omega_{\bar\mu_k}^{\lambda} = [\mathbf{Q}_k^\dagger \mathbf{T}_{kk}^{\lambda}\mathbf{Q}_k]_{\bar\mu\bar\mu}.\label{eq:omegarsp}
\end{equation}
When $\mathbf{T}_{kk}$  has all distinct eigenvalues,
the off-diagonal part of Eq.~(\ref{eq:svdrsp}) leads to 
the OSV relaxation matrix $\mathbf{O}^{\lambda}_k$, expressed in Hadamard product below
\begin{equation}
 \mathbf{O}_k^{\lambda} = 
\Delta\mathbf{G}_k \circ \left [\mathbf{Q}_k^\dagger \mathbf{T}_{kk}^{\lambda}\mathbf{Q}_k \right] 
\label{eq:qrspoff}
\end{equation}
where $\left [\Delta\mathbf{G}_k\right ]_{\bar\mu\bar\nu}
       =\frac{1}{\omega_{\bar\nu_k}-\omega_{\bar\mu_k}}~\mathrm{with~} \bar\mu\neq\bar\nu$.
And the pair-specific relaxation matrix is
\begin{equation}
  \mathbf{O}_{kl}^{\lambda} = 
 \Delta\mathbf{G}_{kl} \circ 
  diag\left(\mathbf{Q}_k^\dagger \mathbf{T}_{kk}^{\lambda}\mathbf{Q}_k,\mathbf{Q}_l^\dagger \mathbf{T}_{ll}^{\lambda}\mathbf{Q}_l\right)
\end{equation}
with
\begin{equation}
 \Delta\mathbf{G}_{kl}=diag\left(\Delta\mathbf{G}_k,\Delta\mathbf{G}_l\right).\label{eq:dgkl}
\end{equation}
Therefore the computation of the off-diagonal element of $\mathbf{O}_k^{\lambda}$ 
requires only the first derivative of $\mathbf{T}_{kk}$ matrix. 

We can prove (c.f. S2 in Supporting Information) that 
the gradient $E_c^\lambda=\frac{\partial E_c}{\partial \lambda}$ 
of OSV-MP2 energy of Eq.~(\ref{eq:ec}) 
is invariant with the rotations among all retained OSVs $\{\bar\mu_k\}$,
\begin{equation}
 \frac{\partial E_c^\lambda}{\partial [\mathbf{O}_k^{\lambda}]_{\bar\mu\bar\nu}}= 0. \label{eq:ecmunu}
\end{equation}
As long as this invariance holds, 
the orbital rotation $\mathbf{O}_k^{\lambda}$ must be made between 
the discarded $\{\bar\mu^\prime\}$ and kept $\{\bar\nu\}$ OSVs
belonging to the subsets of different eigenvalues.  
Therefore the non-degenerate formalism of Eq.~(\ref{eq:qrspoff})
is precisely applicable to $\mathbf{O}_k^{\lambda}$.


The eigenvalue matrix $\mathbf{\Omega}_k = \mathbf{Q}_k^\dagger\mathbf{T}_{kk}\mathbf{Q}_k$ 
can be understood as the projection of the semi-canonical MP2 diagonal amplitude $\mathbf{T}_{kk}$
in the OSV basis, which is diagonal and uniquely defined for each orbital.
As we can show (c.f. S3 in Supporting Information),
the relaxation  $\mathbf{\Omega}_k^\lambda$ must always remain rigorously diagonal as
\begin{equation}
\mathbf{\Omega}_k^\lambda = diag\left (\omega_1^\lambda,\omega_2^\lambda,\cdots\right ).
\label{eq:ptomega}
\end{equation}
With $\mathbf{\Omega}_k(\lambda) = \mathbf{\Omega}_k^0 + \lambda\mathbf{\Omega}_k^\lambda$ 
correct through the first-order expansion, we conclude then that the perturbed
OSV-projected amplitudes $\mathbf{\Omega}_k(\lambda)$ must be diagonal as well
between subspaces belonging to different eigenvalues.  This imposed diagonal
constraint, similar to the canonical condition of Hartree-Fock gradients, has
some convenience, for example,  of allowing in principle different (usually
smaller) OSV gradient domains from original energy domains for more
efficient gradient computation, which will be the subject of our future work.

\subsection{OSV-MP2 analytical gradient theory}

The analytical gradient of the OSV-MP2 correlation energy with respect to a perturbation $\lambda$ 
(eg, an atomic position displacement) can be computed
in terms of  the derivatives of both $\mathbf{K}_{(ij,ij)}^\lambda$ 
and $\mathbf{R}_{(ij,ij)}^\lambda$  in the OSV basis
\begin{equation}
 E^\lambda_c = \frac{dE_c}{d\lambda} =  
\sum_{ij}\mathbf{\langle}\mathbf{K}_{(ij,ij)}^\lambda \overline{\mathbf{T}}_{(ij,ij)}\mathbf{\rangle}
        + \mathbf{\langle}\mathbf{R}_{(ij,ij)}^\lambda \overline{\mathbf{T}}_{(ij,ij)}\mathbf{\rangle},\label{eq:ecgrad1}
\end{equation}
whereas the amplitudes $\overline{\mathbf{T}}_{(ij,ij)}$
make no contribution as they are simply variational to $E_c$.
It is obvious that the derivatives $\mathbf{K}_{(ij,ij)}^\lambda$ and 
$\mathbf{R}_{(ij,ij)}^\lambda$ must be jointly determined through 
the responses of the OSVs, LMOs and AOs.
The MP2 energy gradient of Eq.~(\ref{eq:ecgrad1}) thus  consists of
the relaxation contributions from
OSVs ($E^{\{\lambda\}}_c$), MOs ($ E^{[\lambda]}_c$) and AOs ($E^{(\lambda)}_c$), 
respectively,
\begin{equation}
  E^\lambda_c =  E^{\{\lambda\}}_c + E^{[\lambda]}_c + E^{(\lambda)}_c .
\end{equation}

\subsubsection{OSV-specific energy gradient $E^{\{\lambda\}}_c$}

The OSV-specific energy gradient  $E_c^{\{\lambda\}}$ 
is determined by
\begin{equation}
 E^{\{\lambda\}}_c  = 
 4\sum_{ij} \mathbf{\langle}
 \left[
  (\overline{\mathbf{T}}_{(ij,ij)}\mathbf{K}_{(ij,ij)}
   + \mathbf{M}_{ij})
\mathbf{O}_{ij}^{\lambda} 
 \right] \mathbf{\rangle}
\label{eq:ecosvgrad}
\end{equation}
which requires the OSV derivatives of the quantities $\mathbf{A}_{(ij,kl)}$ 
associated with $(i,k)$, $(i,l)$, $(j,k)$ and $(j,l)$ pairs,
such as the exchange integral $\mathbf{K}_{(ij,kl)}$, 
overlap $\mathbf{S}_{(ij,kl)}$ and
the OSV block of the Fock matrix $\mathbf{F}_{(ij,kl)}$,
according to Eq.~(\ref{eq:aijklgrad}).
Here the intermediate $\mathbf{M}_{ij}$ is specific to the pair $ij$, 
arising from the residual contribution $\mathbf{R}_{(ij,ij)}$
in the second term of Eq.~(\ref{eq:rij}),
\begin{equation}
\mathbf{M}_{ij} = \mathbf{D}_{(ij,ij)}\mathbf{F}_{(ij,ij)} + \mathbf{D}_{(ij,ij)}^\prime \mathbf{S}_{(ij,ij)}
- \sum_k \left[f_{jk}\mathbf{D}_{(ij,ik)}\mathbf{S}_{(ik,ij)}+f_{ik}\mathbf{D}_{(ij,kj)}\mathbf{S}_{(kj,ij)}\right],
\end{equation}
The OSV-OSV blocks of the unrelaxed overlap- and energy-weighted density matrices 
are hermitian and defined as $\mathbf{D}_{(ij,kl)}$ and $\mathbf{D}_{(ij,kl)}^\prime$, respectively,
\begin{equation}
  \mathbf{D}_{(ij,kl)}  =  \frac{1}{2}\left[\overline{\mathbf{T}}_{(ij,ij)}\mathbf{S}_{(ij,kl)}\mathbf{T}_{(kl,kl)} 
  + \overline{\mathbf{T}}_{(ij,ij)}^\dagger\mathbf{S}_{(ij,kl)}\mathbf{T}_{(kl,kl)}^\dagger \right ]    \label{eq:osvdm}
\end{equation}
\begin{equation}
  \mathbf{D}_{(ij,kl)}^\prime  =  \frac{1}{2}\left[\overline{\mathbf{T}}_{(ij,ij)}\mathbf{F}_{(ij,kl)}\mathbf{T}_{(kl,kl)} 
  + \overline{\mathbf{T}}_{(ij,ij)}^\dagger\mathbf{F}_{(ij,kl)}\mathbf{T}_{(kl,kl)}^\dagger \right ] \label{eq:osvdmp}
\end{equation}
Since the gradients $E_c^{\{\lambda\}}$ are invariant with the rotations among all kept OSVs $\{\bar\mu_k\}$,
$\mathbf{K}_{(ij,ij)}$ and $\mathbf{M}_{ij}$ 
must involve the discarded OSVs at the dimensions attached to $\mathbf{O}_{ij}^{\{\lambda\}}$,
while the amplitudes $\mathbf{T}_{(ij,ij)}$ and $\overline{\mathbf{T}}_{(ij,ij)}$
remain  within the retained OSV subspace.

\begin{figure}
\centering
\includegraphics[scale=0.50]{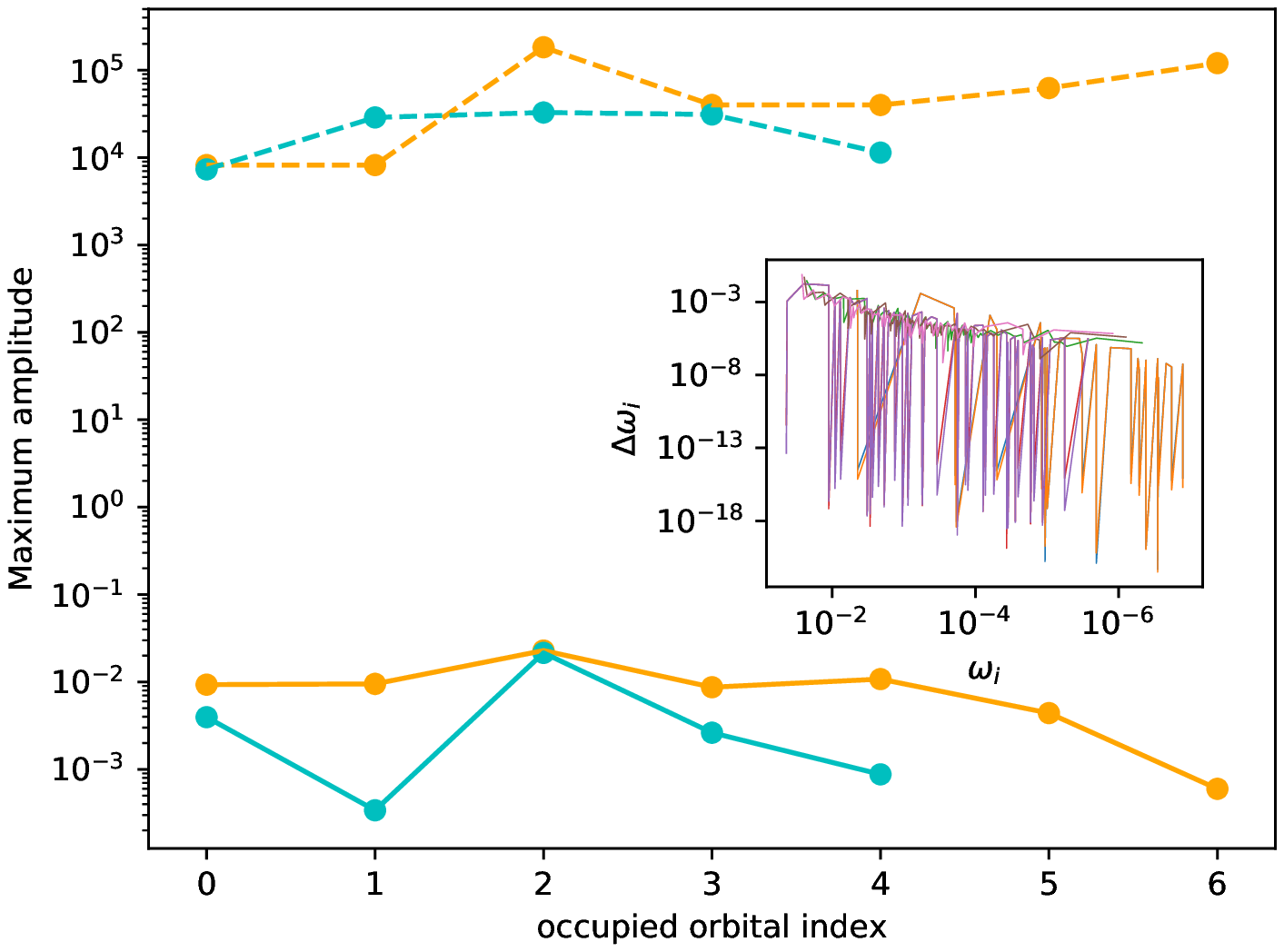} 
\includegraphics[scale=0.50]{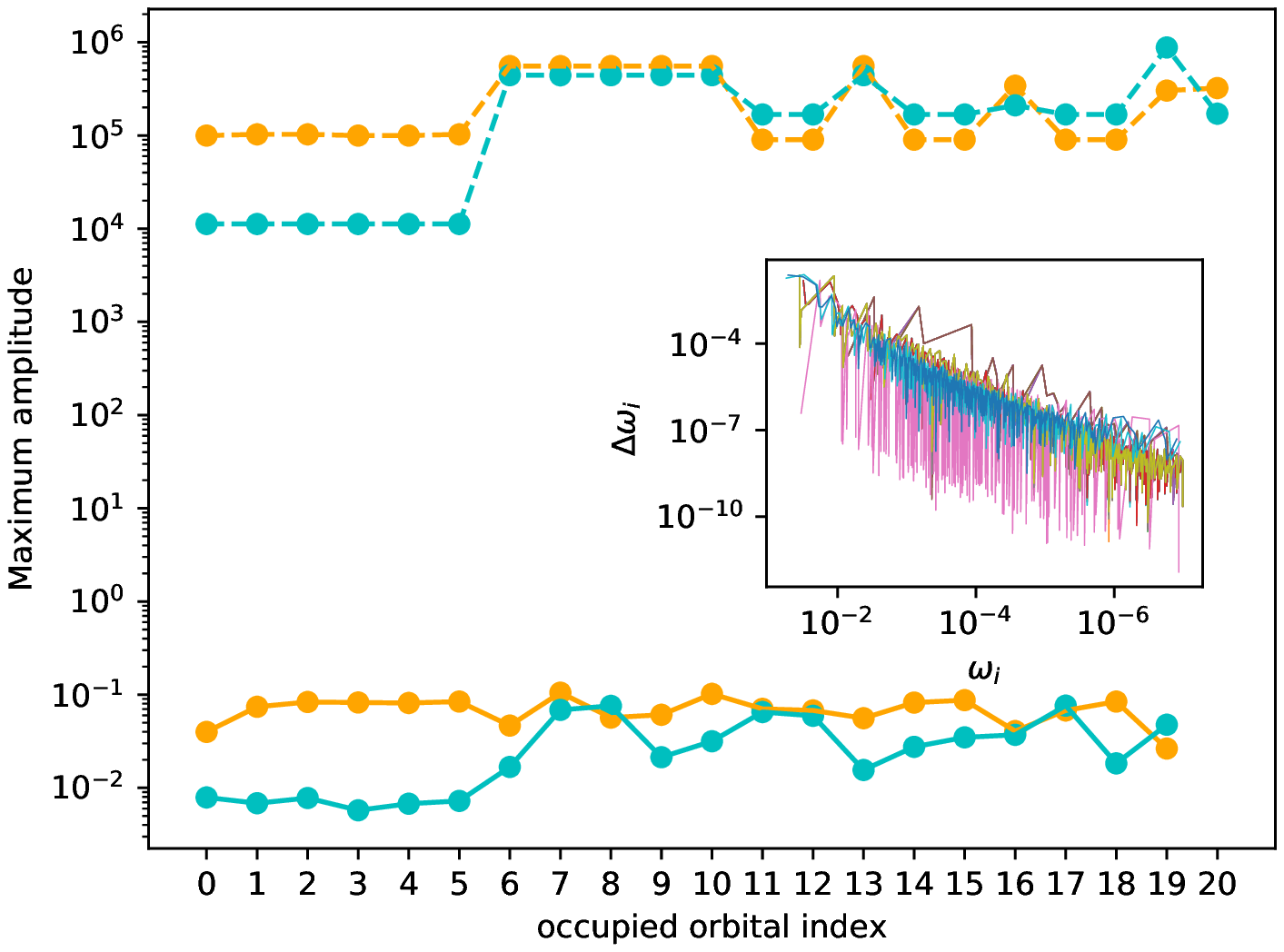} 
\caption{The maximum elements of the Hadamard product 
$\mathbf{N}_{ij} \circ \Delta\mathbf{G}_{ij}^\mathrm{T}$
(solid dot) and $\Delta\mathbf{G}_{ij}$ (dashed dot) 
computed with cc-pvtz (cyan) and aug-cc-pvtz (orange) with respect to the indices of occupied orbitals 
for N$_2$ (left) and C$_6$H$_6$ (right). 
The insets show the magnitudes of $\omega_{\bar\mu_i}-\omega_{\bar\nu_i^\prime}$ 
with respect to all eigenvalues, with each solid curve specific to one orbital.}\label{fig:n2c6h6}
\end{figure}
The evaluation of $\mathbf{O}^{\lambda}_{ij}$
employs the non-degenerate formalism using all pairs of distinct eigenvalues 
associated with the discarded and retained OSV subspaces, respectively.
Substituting Eq.~(\ref{eq:qrspoff}), $E_c^{\{\lambda\}}$ can be rewritten as

\begin{equation}
 E_c^{\{\lambda\}} = 4\sum_{ij} \mathbf{\langle}
  \left( \mathbf{N}_{ij} \circ \Delta\mathbf{G}_{ij}^\mathrm{T} \right)
 diag\left(\mathbf{Q}_i^\dagger \mathbf{T}_{ii}^{\lambda}\mathbf{Q}_i,\mathbf{Q}_j^\dagger \mathbf{T}_{jj}^{\lambda}\mathbf{Q}_j\right)
\mathbf{\rangle}
\end{equation}
with 
\begin{equation}
\mathbf{N}_{ij} = \overline{\mathbf{T}}_{(ij,ij)}\mathbf{K}_{(ij,ij)} + \mathbf{M}_{ij}
\label{eq:nij}
\end{equation}
and $\Delta\mathbf{G}_{ij}^\mathrm{T}$ is given in Eq.~(\ref{eq:dgkl}).
The numerical stability of computed  $E_c^{\{\lambda\}}$ 
can be demonstrated by illustrating the maximum element of 
$\mathbf{N}_{ij} \circ \Delta\mathbf{G}_{ij}^\mathrm{T}$
for each orbital in Figure~\ref{fig:n2c6h6}.
To this end, we choose N$_2$ and C$_6$H$_6$ which own high symmetry and thus
a larger number of near-degenerate eigenvalues of the semi-canonical MP2 diagonal amplitudes. 
As seen in the insets of Figure~\ref{fig:n2c6h6},
it is evident that the large values of $\frac{1}{\Delta\omega_i}$
due to the vanishingly small difference $\Delta\omega_i=\omega_{\bar\mu_i}-\omega_{\bar\nu_i^\prime}$ 
are largely compensated by $\mathbf{N}_{ij}$,
which in fact yields smooth analytical gradients,
without instability hurdles in practice.

\subsubsection{MO-specific energy gradient $E^{[\lambda]}_c$} \label{sec:mogra}

The sparse structure of the OSV-MP2 amplitudes is most favorably exploited
with the locality of LMOs.
The occupied canonical MOs are localized using  Pipek-Mezey (PM)\cite{pipek1989fast} with 
meta-L\"owdin atomic charges for their good transferability in  different molecular environments
created by the variation of atomic positions\cite{sun2014exact}. 
For evaluating the meta-L\"owdin charges, the core and valence orbitals
are distinguished  based on the locality of the predefined NAO (natural atomic orbital),
and then L\"owdin-orthogonalized within their own space.
The localization procedure introduces a new transformation matrix 
$\mathbf{L}(\lambda)$
that transforms the occupied canonical MOs $\mathbf{C}_o(\lambda)$ 
into the orthonormal LMOs $\mathbf{\tilde{C}}_o(\lambda)$,
which must hold as well for a system under the perturbation $\lambda$,
\begin{equation}
 \mathbf{\tilde{C}}_o(\lambda) = \mathbf{C}_o(\lambda)\mathbf{L}(\lambda)
 \label{eq:lmo}
\end{equation}
with the orthonormal condition 
$\mathbf{L}^\dagger(\lambda)\mathbf{L}(\lambda) = \mathbf{1}$.
The LMO response $\mathbf{\tilde{C}}_o^{[\lambda]}$
therefore arises from both derivative contributions 
of $\mathbf{U}^{[\lambda]}$ and $\mathbf{\tilde{U}}^{[\lambda]}$,
\begin{equation}
 \mathbf{\tilde{C}}^{[\lambda]}_o= \mathbf{C}^0\mathbf{U}^{[\lambda]}\mathbf{L}^0
  + \mathbf{\tilde{C}}^0_o\mathbf{\tilde{U}}^{[\lambda]}.
 \label{eq:lmograd}
\end{equation}
The coupled-perturbed localization (CPL) described in Ref.\cite{el1998analytical}
for PM  localization function
and the coupled-perturbed Hartree-Fock equations are solved
to determine $\mathbf{\tilde{U}}^{[\lambda]}$ and $\mathbf{U}^{[\lambda]}$, respectively.
However, neither $\mathbf{\tilde{U}}^{[\lambda]}$ nor
$\mathbf{U}^{[\lambda]}$ is explicitly computed or stored in our implementation
for reasons of computational efficiency,
and their contributions are merged into 
the OSV-based Z-vector equation.

As seen in Eqs. (\ref{eq:rij}) and (\ref{eq:ec}), 
apparently the MO-specific $E^{[\lambda]}_c$ is determined  by
the quantities that involve the derivatives with respect to LMOs and canonical virtual MOs,
i.e., the derivatives of the  exchange integral and the Fock matrix,
\begin{equation}
 E^{[\lambda]}_c = 
  2\sum_{ij}
  \mathbf{\langle}\mathbf{K}_{(ij,ij)}^{[\lambda]}
\overline{\mathbf{T}}_{(ij,ij)} \mathbf{\rangle}
+ \mathbf{\langle}\mathbf{D}_{(ij,ij)}
  \mathbf{F}_{(ij,ij)}^{[\lambda]}\mathbf{\rangle}
- 2D_{ij} f_{ij^{[\lambda]}}.
\label{eq:ecgrad2}
\end{equation}
The occupied-occupied elements of the unrelaxed density matrix is
\begin{equation}
 D_{ij} = \frac{1}{2}\sum_k \mathbf{\langle}\mathbf{T}_{(ki,ki)}\mathbf{S}_{(ki,kj)}
\overline{\mathbf{T}}_{(kj,kj)}\mathbf{S}_{(kj,ki)} 
+ \overline{\mathbf{T}}_{(ki,ki)}\mathbf{S}_{(ki,kj)}
\mathbf{T}_{(kj,kj)}\mathbf{S}_{(kj,ki)} 
\mathbf{\rangle}.
\label{eq:occdm}
\end{equation}
According to Eq.~(\ref{eq:aijkl}), we have
\begin{equation}
\mathbf{K}_{(ij,ij)}^{[\lambda]}  = 
 \left( 
  {\begin{array}{c} 
    \mathbf{Q}_{i}^\dagger\\\mathbf{Q}_{j}^\dagger
   \end{array}}
 \right) 
\mathbf{K}_{ij}^{[\lambda]}
 \left( 
  {\begin{array}{cc} 
    \mathbf{Q}_{i} & \mathbf{Q}_{j}
   \end{array}}
 \right),~ 
\mathbf{F}_{(ij,ij)}^{[\lambda]}  =  
 \left( 
  {\begin{array}{c} 
    \mathbf{Q}_{i}^\dagger\\\mathbf{Q}_{j}^\dagger
   \end{array}}
 \right) 
\mathbf{F}^{[\lambda]}
 \left( 
  {\begin{array}{cc} 
    \mathbf{Q}_{i} & \mathbf{Q}_{j}
   \end{array}}
 \right) \label{eq:kijijgradmo}
\end{equation}
where
\begin{equation}
\mathbf{K}_{ij}^{[\lambda]}  = 
\mathbf{K}_{ij^{[\lambda]}} + \mathbf{K}_{i^{[\lambda]}j}+\mathbf{K}_{ij}^{0[\lambda]}+\mathbf{K}_{ij}^{[\lambda]0},~ 
\mathbf{F}^{[\lambda]}  =  \mathbf{F}^{0[\lambda]}+\mathbf{F}^{[\lambda]0} \label{eq:fgradmo} 
\end{equation}
with superscripts $[\lambda]$ for the MO derivatives.
Substituting Eqs. (\ref{eq:kijijgradmo})--(\ref{eq:fgradmo}) 
into Eq.~(\ref{eq:ecgrad2}) and utilizing the particle permutation symmetry, 
we arrive at the MO-specific energy gradient
\begin{eqnarray}
 E^{[\lambda]}_c &=& 
 4 \sum_{ij}
   \mathbf{\langle}
\overline{\mathbf{T}}_{(ij,ij)} 
 \left( 
  {\begin{array}{c} 
    \mathbf{Q}_{i}^\dagger\\\mathbf{Q}_{j}^\dagger
   \end{array}}
 \right)
\left(\mathbf{K}_{ij^{[\lambda]}} + \mathbf{K}_{ij}^{0[\lambda]}\right)
 \left( 
  {\begin{array}{cc} 
    \mathbf{Q}_{i} & \mathbf{Q}_{j} 
   \end{array}}
 \right) 
\mathbf{\rangle}\nonumber \\ 
 & & +
 \mathbf{\langle}\mathbf{D}_{(ij,ij)}
 \left( 
  {\begin{array}{c} 
    \mathbf{Q}_{i}^\dagger\\\mathbf{Q}_{j}^\dagger
   \end{array}}
 \right)
\mathbf{F}^{0[\lambda]}
 \left( 
  {\begin{array}{cc} 
    \mathbf{Q}_{i} & \mathbf{Q}_{j} 
   \end{array}}
 \right) 
\mathbf{\rangle}
  -  D_{ij}f_{ij^{[\lambda]}}.
\label{eq:ecmograd}
\end{eqnarray}
Above, 
$\mathbf{F}^{0[\lambda]}$ and 
$\mathbf{K}_{ij}^{0[\lambda]}$ are associated with the relaxation of 
one virtual MO,
\begin{equation}
[\mathbf{F}^{0[\lambda]}]_{ab} =  
f_{aa}U^{[\lambda]}_{ab},
~[\mathbf{K}_{ij}^{0[\lambda]}]_{ab}=
 \sum_p(ia \rvert jp)U^{[\lambda]}_{pb}.
\end{equation}
$\mathbf{K}_{ij^{[\lambda]}}$ 
and $f_{ij^{[\lambda]}}$ are the derivatives
with respect to one of the LMOs, respectively,
which can be evaluated according to Eq.~(\ref{eq:lmograd}),
\begin{equation}
\mathbf{K}_{ij^{[\lambda]}} = 
\sum_{al} \mathbf{K}_{ia}
               U^{[\lambda]}_{aj}
+ \sum_{kl} \mathbf{K}_{ik} 
   (U^{[\lambda]}_{kj}
   + \tilde{U}^{[\lambda]}_{kj})
 \label{eq:kijgraduv}
\end{equation}
\begin{equation}
f_{ij^{[\lambda]}} = 
\sum_{k} f_{ik}
(U^{[\lambda]}_{kj} + \tilde{U}^{[\lambda]}_{kj}).
 \label{eq:fijgraduv}
\end{equation}
In Eqs. (\ref{eq:kijgraduv}) and (\ref{eq:fijgraduv}),
the symmetric block of $\mathbf{U}^{[\lambda]}$ is transformed into LMOs,
and depends solely on the AO derivative of overlap matrix
according to the MO orthonormal condition.
However, the off-diagonal block of $\mathbf{U}^{[\lambda]}$
accounts for the rotation of MOs between the occupied and virtual spaces,
which is solved in the OSV Z-vector approach.

\subsubsection{AO-specific energy gradient $E^{(\lambda)}_c$}\label{sec:aogra}

$E^{(\lambda)}_c$ simply evaluates the energy expression of Eq.~(\ref{eq:ec})
in terms of AO derivative integrals, 
the occupied-occupied block (Eq.~(\ref{eq:occdm})) and 
OSV-OSV block (Eq.~(\ref{eq:osvdm})) of the unrelaxed density matrices,
\begin{equation}
 E^{(\lambda)}_c = 
  2\sum_{ij}
  \mathbf{\langle}\overline{\mathbf{T}}_{(ij,ij)}\mathbf{K}_{(ij,ij)}^{(\lambda)}\mathbf{\rangle}
+ \mathbf{\langle}\mathbf{D}_{(ij,ij)}\mathbf{F}_{(ij,ij)}^{(\lambda)}\mathbf{\rangle}
- D_{ij} f_{ij}^{(\lambda)}.
\label{eq:ecaograd}
\end{equation}
The OSV overlap $\mathbf{S}_{(ij,kl)}=\left({\begin{array}{c}\mathbf{Q}_{i}^\dagger \\ \mathbf{Q}_{j}^\dagger\end{array}}\right)
\left({\begin{array}{cc}\mathbf{Q}_{k}&\mathbf{Q}_{l}\end{array}}\right)$
makes no contribution here to the AO-specific energy gradient.
The corresponding two- and one-electron derivative integrals for an $ij$ pair
are computed using their AO derivative integrals, 
including the AO derivatives of 
the exchange integral matrix $\mathbf{K}_{(ij,ij)}^{(\lambda)}$, 
the OSV-OSV block of the Fock matrix $\mathbf{F}_{(ij,ij)}^{(\lambda)}$,
and the occupied-occupied Fock elements $f_{ij}^{(\lambda)}$.

\subsection{Implementation scheme}
Computing the OSV-, MO- and AO-specific two-electron contributions to 
the OSV-MP2 energy gradient according to Eqs.~(\ref{eq:ecosvgrad}), (\ref{eq:ecmograd}) 
and (\ref{eq:ecaograd}) would be straightforward with
yet unfortunately very demanding expenses.
The primary bottleneck originates from the evaluation and transformation of 
the subsumed exchange integral $\mathbf{K}_{(ij,ij)}$ and the 
AO/MO derivatives $\mathbf{K}_{ij}^{(\lambda)}$, $\mathbf{K}_{ij^{[\lambda]}}$
and $\mathbf{K}_{ij}^{0[\lambda]}$ involving more than two virtual MO indices.
Both computational storage and operation costs  increase rapidly 
with sizes of molecule.
Significant savings can be achieved by employing the resolution of identity (RI) 
technique~\cite{feyereisen1993use,weigend1998ri}.
In the present work, RI approximate exchange integrals and their derivatives  
are implemented  in adaption to OSV basis
for accelerated evaluation and transformation.
According to the RI scheme in the Coulomb metric, 
the four-center two-electron (4c2e) integral
$(ip\vert jq)$ is approximated as a simple product 
of the lower-rank three-center two-electron (3c2e) integrals 
$\mathbf{J}_i$ and $\mathbf{J}_j$, specific to each LMO $i$ and $j$, respectively, 
\begin{equation}
(ip\rvert jq) 
      = \left[\mathbf{J}_i^\dagger \mathbf{J}_j\right]_{pq}
\label{eq:3c2e}
\end{equation}
with the 3c2e matrix element $[\mathbf{J}_i]_{Ap}=\sum_B[\mathbf{V}^{-\frac{1}{2}}]_{AB}(B\rvert i p)$
in terms of a set of auxiliary basis functions $\{A,B,\cdots\}$,
and $\mathbf{V}$ denotes  the Coulomb metric matrix
\begin{equation}
 [\mathbf{V}]_{AB} = \iint d\vec{r}_1 d\vec{r}_2 \frac{A(\vec{r}_1)B(\vec{r}_2)}{|\vec{r}_1-\vec{r}_2|}.
\end{equation}
In the following, we use $\mathbf{J}_i^o$ and $\mathbf{J}_i^v$ 
for the occupied $p=j$ and virtual $p=a$ blocks, respectively.

In our OSV-MP2 gradient formulation, we must however  deal with the integrals
$(i\bar\mu_j|A)$ and $(i\bar\mu_j^\prime|A)$ in both kept and discarded OSV basis for treating OSV relaxation.
The number of these integrals for all $(i,j)$ pairs
grows as $\mathcal{O}(O^2VN_{aux})$,
and the storage becomes rather unfavorable for large molecules 
if they are explicitly computed.
To avoid such high storage costs, we have exploited an implementation
in which the 3c2e MO integrals $\mathbf{J}_i$ are transformed into an intermediate $\mathbf{Y}_i$
accounting for two-electron contributions to the OSV-MP2 gradient from both MO and OSV rotations,
\begin{equation}  \label{eq:ec2eY}
  \mathbf{Y}_i = 
  \sum_{j}  \mathbf{J}_{j}^v\left(\mathbf{Q}_i~\mathbf{Q}_j \right)\
  \overline{\mathbf{T}}_{(ij,ij)}
  \left(
  \begin{array}{c}
    \mathbf{Q}_i^\dagger \\
    \mathbf{Q}_j^\dagger
  \end{array}
  \right)
 + \mathbf{J}_i^v(\mathbf{X}_{ij}^\top+\mathbf{X}_{ji}^\bot)  
\end{equation}
where
\begin{equation}
\left[\mathbf{X}_{ij}^{\top} \right]_{ab} = \frac{2\left[\mathbf{Q}_i\left(\mathbf{N}_{ij}\circ\Delta\mathbf{G}_{ij}^\mathrm{T}\right)^{\top}\mathbf{Q}_i^{\prime\dagger} \right]_{ab}}{f_{aa}+f_{bb}-2f_{ii}},~
\left[\mathbf{X}_{ij}^{\bot} \right]_{ab} = \frac{2\left[\mathbf{Q}_j\left(\mathbf{N}_{ij}\circ\Delta\mathbf{G}_{ij}^\mathrm{T}\right)^{\bot}\mathbf{Q}_j^{\prime\dagger} \right]_{ab}}{f_{aa}+f_{bb}-2f_{jj}}
\end{equation}
with the symbols $\top$ and $\bot$ denoting the upper and lower diagonal blocks.
$\mathbf{X}_{ij}^{\top,~\bot}$ are computed and accessed on the fly for each $(i,j)$ pair.
The one-index transformations made in $\mathbf{X}_{ij}$ are carried out with the kept ($\mathbf{Q}_i$)
and discarded ($\mathbf{Q}_i^{\prime\dagger}$) OSV orbitals.
Both $\mathbf{J}_i$, AO derivative $\mathbf{J}_i^{(\lambda)}$ and $\mathbf{Y}_i$
are of the row dimension $N_{aux}$ and column dimension $V$, 
and  can be conveniently stored on disk as their total number grows as $\mathcal{O}(OVN_{aux})$,
forming no major obstacle for a usual range of molecular sizes.
In our implementation, the dominant formal operation scales as $\mathcal{O}(O^2N_{osv}VN_{aux})$ 
for computing $\mathbf{Y}_i$ and $\mathcal{O}(ON_{osv}N_{osv}^\prime N_{aux})$ for $\mathbf{X}_{ij}$,
where $N_{osv}$ and $N_{osv}^\prime$ are the number of the kept and discarded OSVs, respectively.
Nevertheless, when working with reasonably selected OSVs and pairs for a good accuracy-cost balance,
the actual computational  cost can be reduced to $\mathcal{O}(N^{3\sim 4})$.

By combining $E_c^{(\lambda)}$, $E_{c}^{[\lambda]}$ and $E_{c}^{\{\lambda\}}$,
our working equation for evaluating the OSV-MP2 energy gradient can be written 
in terms of the AO-derivatives of Fock ($\mathbf{F}^{(\lambda)}$ and $f_{ij}^{(\lambda)}$),
overlap ($\mathbf{S}^{(\lambda)}$ and $S_{ij}^{(\lambda)}$) 
and 3c2e integral ($\mathbf{J}_i^{v(\lambda)}$) matrices,
\begin{eqnarray}
 E_c^{\lambda}
&=& 2\mathbf{\langle}(\mathbf{D}+\overline{\mathbf{D}})\mathbf{F}^{(\lambda)}\mathbf{\rangle}
  - 2\sum_{ij}(D_{ij}+\delta_{ij}\overline{D}_{ii})f_{ij}^{(\lambda)}
-  2\mathbf{\langle}(\mathbf{D}^{\prime}+\overline{\mathbf{D}}^\prime
+ \sum_i \mathbf{J}_i^{v\dagger}\mathbf{Y}_i)\mathbf{S}^{(\lambda)}\mathbf{\rangle}
  \nonumber \\
& &  
+ 2\sum_{\alpha\beta} (\sum_{ij}\tilde{C}_{\alpha i}\Gamma_{ij}\tilde{C}_{\beta j}
                           -\Lambda_{ij}\mathcal{A}_{ij,\alpha\beta}) S_{\alpha\beta}^{(\lambda)}
+ 4\mathbf{\langle} \sum_i \mathbf{Y}_i^\dagger(\mathbf{J}_i^{v(\lambda)}+\mathbf{J}_i^o\mathbf{S}_{ov}^{(\lambda)})\mathbf{\rangle}
+ 4\mathbf{\langle} \mathbf{Z}^\dagger\mathbf{B}^{(\lambda)}\mathbf{\rangle}.\nonumber \\ 
 \label{eq:ec1e2e}
\end{eqnarray}
The unrelaxed ($\mathbf{D}$) and relaxed ($\overline{\mathbf{D}}$) density matrices 
are utilized in MO basis,
\begin{eqnarray}
 \mathbf{D} & = & \sum_{ij}\left( \mathbf{Q}_i~\mathbf{Q}_j\right) 
\mathbf{D}_{(ij,ij)}  
 \left(
  \begin{array}{c}
    \mathbf{Q}_i^\dagger \\
    \mathbf{Q}_j^\dagger
  \end{array}
 \right), \\
\overline{\mathbf{D}} & = & \sum_{ij} 
\mathbf{T}_{ii}\left(\mathbf{X}_{ij}^\top + \mathbf{X}_{ji}^\bot \right),\\
\overline{D}_{ii}& = & \mathbf{\langle}
\mathbf{T}_{ii}\sum_j\left(\mathbf{X}_{ij}^\top + \mathbf{X}_{ji}^\bot \right)\mathbf{\rangle},
\end{eqnarray}
and the energy-weighted unrelaxed ($\mathbf{D}^\prime$) and relaxed ($\overline{\mathbf{D}}^\prime$) 
density matrices are,
\begin{equation}
\left[\mathbf{D}^{\prime}\right]_{ab} = \frac{1}{2}(f_{aa}+f_{bb})\left[\mathbf{D}\right]_{ab},
~D^{\prime}_{ij} = \sum_k f_{ik}D_{kj},
\end{equation}
\begin{equation}
\left[\overline{\mathbf{D}}^\prime \right]_{ab} =\frac{1}{2}(f_{aa}+f_{bb})\left[\overline{\mathbf{D}}\right]_{ab}, 
~\overline{D}_{ij}^\prime=f_{ij}\overline{D}_{ii}
\end{equation}
The fourth term needs the $\mathbf{\Gamma}$ matrix,
\begin{equation}
\Gamma_{ij}= D_{ij}^{\prime}+\overline{D}_{ij}^\prime  + \mathbf{\langle}\mathbf{J}^{v\dagger}_i \mathbf{Y}_j\mathbf{\rangle}
\end{equation}
as well as $\mathbf{\Lambda}$ matrix that are obtained by solving the following linear CPL equation 
for PM localization constraint,
\begin{equation}
 \mathbf{\mathcal{C}}^\dagger \mathbf{\Lambda} = \mathbf{\Gamma}^\dagger.
\end{equation}
Finally, $\mathbf{B}^{(\lambda)}$ of the last term in Eq.~(\ref{eq:ec1e2e})
collects all AO-derivatives in the Fock and overlap matrices, compuated only once and for all, 
\begin{equation}
 [\mathbf{B}^{(\lambda)}]_{ai} = -[\mathbf{F}^{(\lambda)}]_{ai} + [\mathbf{S}^{(\lambda)}]_{ai}
    +\frac{1}{2}\sum_{kl}[\mathbf{A}]_{ai,kl}S^{(\lambda)}_{kl} 
\end{equation} 
where
\begin{equation}
[\mathbf{A}]_{ai,bj}=\delta_{ab}\delta_{ij}(f_{aa}-f_{ii})+4(ai\rvert jb)-(ab\rvert ij)-(aj\rvert ib)
\end{equation}
for which the two-electron integrals are evaluated with  RI approximation.
The remaining Z-vector $\mathbf{Z}$ must be solved in the other linear equation
\begin{equation} \label{eq:zvec}
 \mathbf{A}^\dagger \mathbf{Z} = \mathbf{W}
\end{equation}
The source term takes the form below,
\begin{equation}
[\mathbf{W}]_{ai}= \mathbf{\langle}\mathbf{J}_a^v\mathbf{Y}_i^{\dagger}\mathbf{\rangle} + \sum_j[\mathbf{Y}_j^\dagger\mathbf{J}_j^o]_{ai}
+ 2\sum_{kl}{\Lambda}_{kl}\mathcal{B}_{kl,ai}
\end{equation}
Finally, the explicit mathematical forms of the intermediates 
$\mathcal{A}_{ij,\alpha\beta}$, $\mathcal{B}_{kl,ai}$ and $\mathcal{C}_{kl,ij}$
are specified in Eqs. (28)--(30) in Ref.\cite{el1998analytical}, and thus will not be repeated here.

\section{APPLICATIONS TO MOLECULAR STRUCTURES}

\subsection{Accuracy of OSV-MP2 analytical gradients}

The correctness of our implementation has been examined 
by comparing the OSV-MP2 analytical gradients with 
OSV-MP2 numerical gradients for N$_2$ and water clusters (H$_2$O)$_n$ ($n=1-3$).
The root mean square deviations (RMSDs) of the gradient differences are about
$10^{-6}$--$10^{-7}$ a.u. for various OSV selections 
($l_{osv}=10^{-3}$, $10^{-4}$ and $10^{-7}$).

\begin{figure}
 \begin{minipage}{8cm}
  \centering
  \includegraphics[width=8cm]{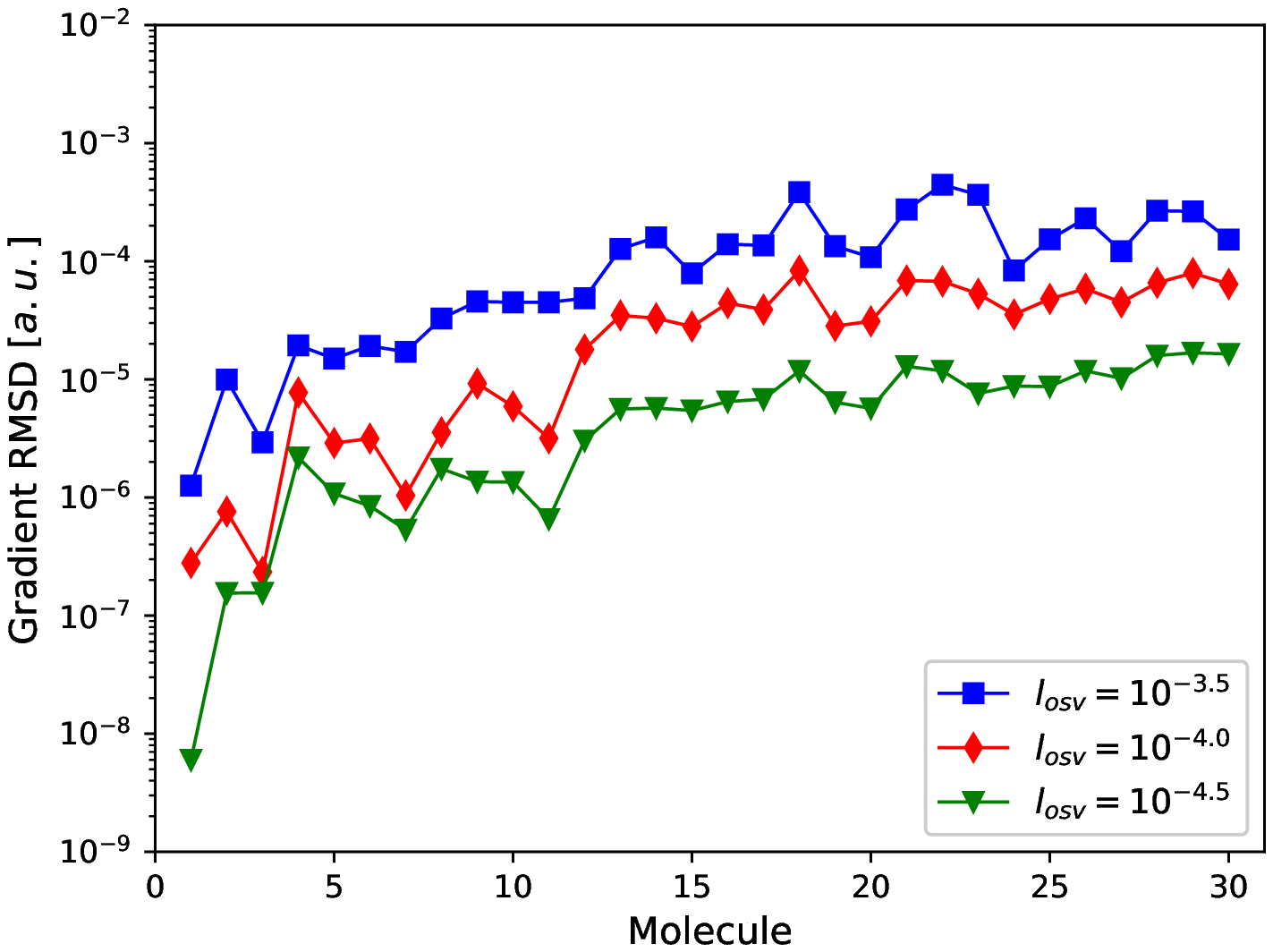}
  \caption*{(a)}
 \end{minipage}
 \begin{minipage}{8cm}
  \centering
  \includegraphics[width=8cm]{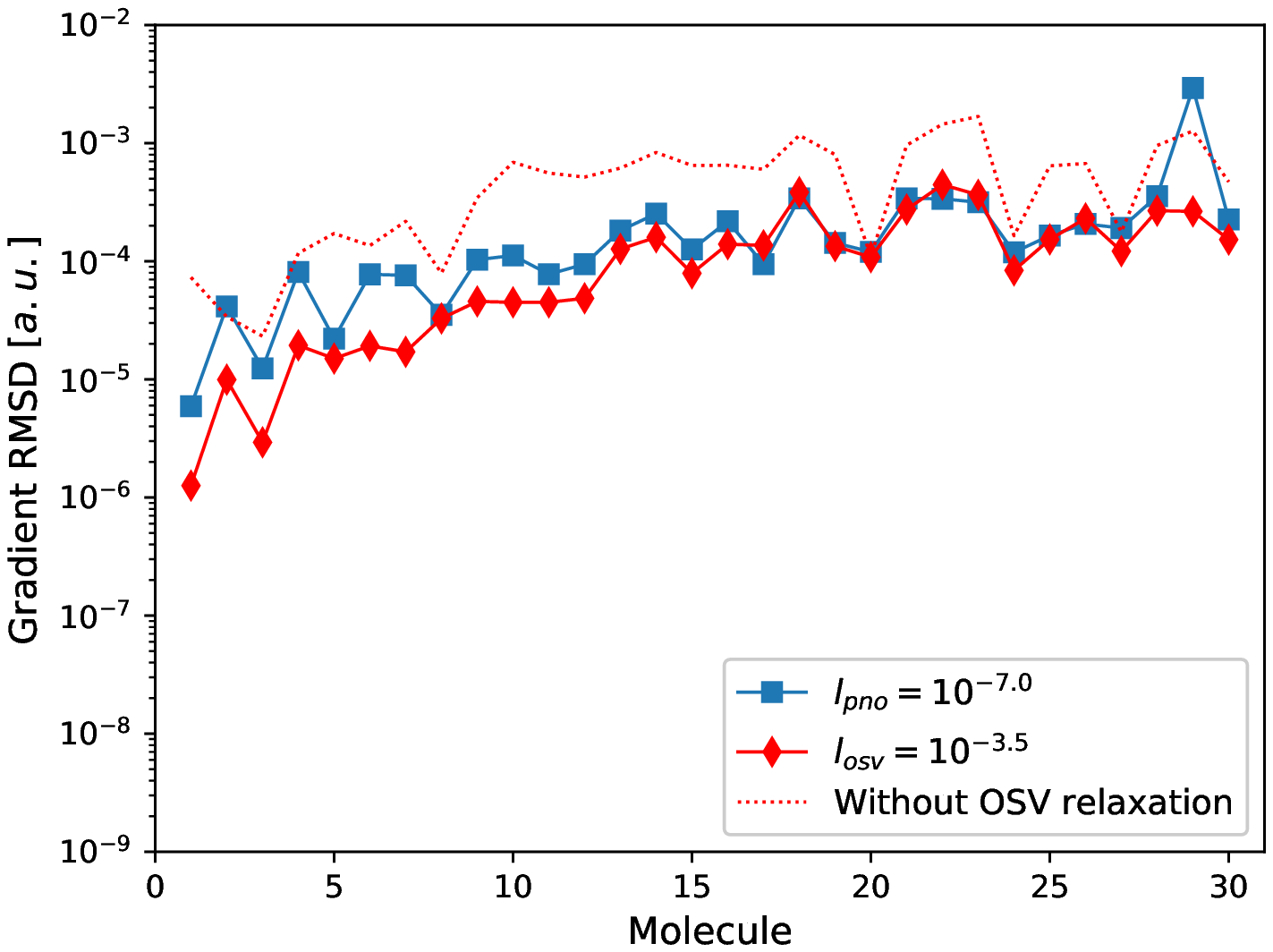} 
  \caption*{(b)}
 \end{minipage}%
 \\
 \begin{minipage}{8cm}
  \centering
  \includegraphics[width=8cm]{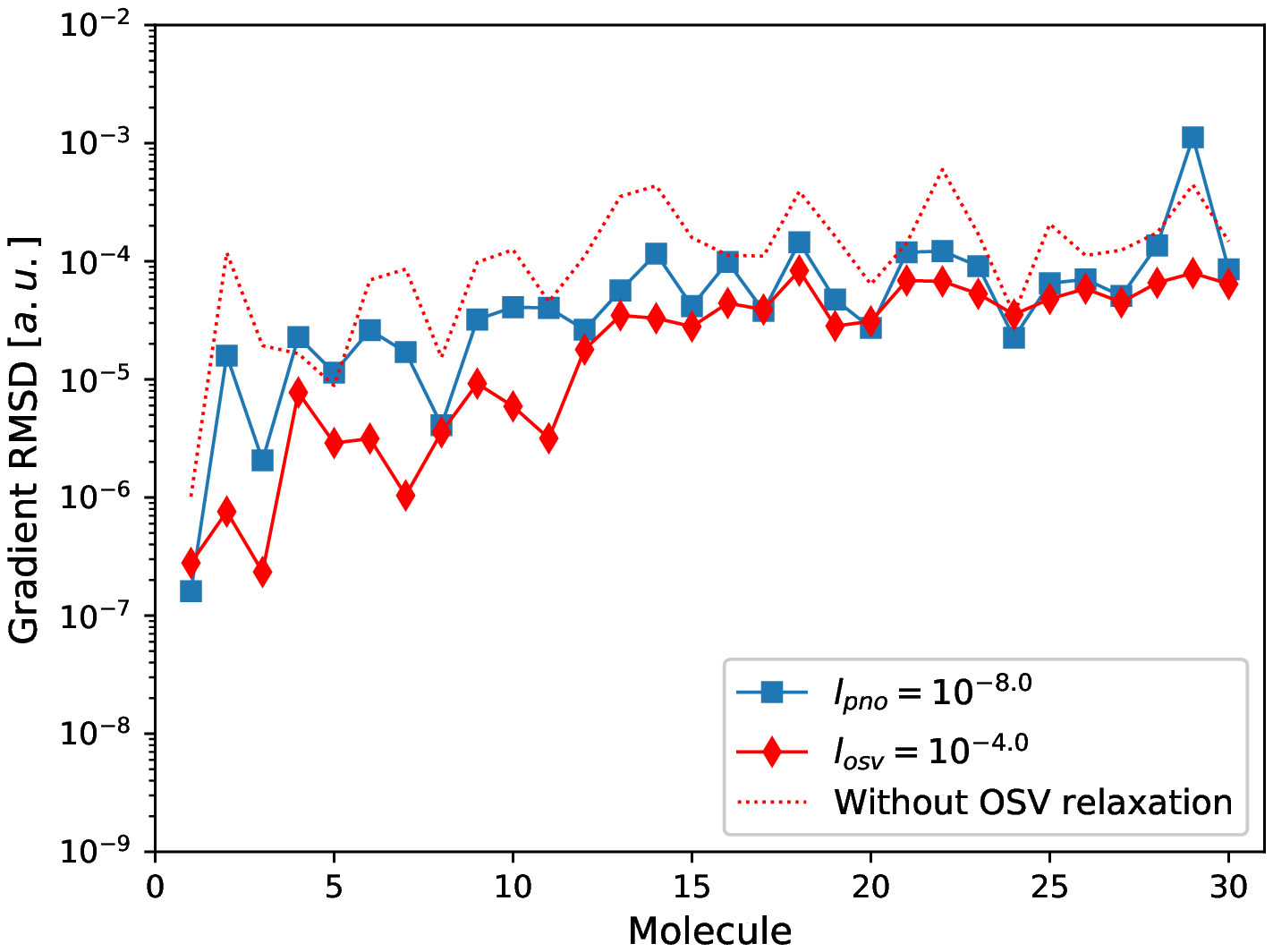}
  \caption*{(c)}
 \end{minipage}
 \begin{minipage}{8cm}
  \centering
  \includegraphics[width=8cm]{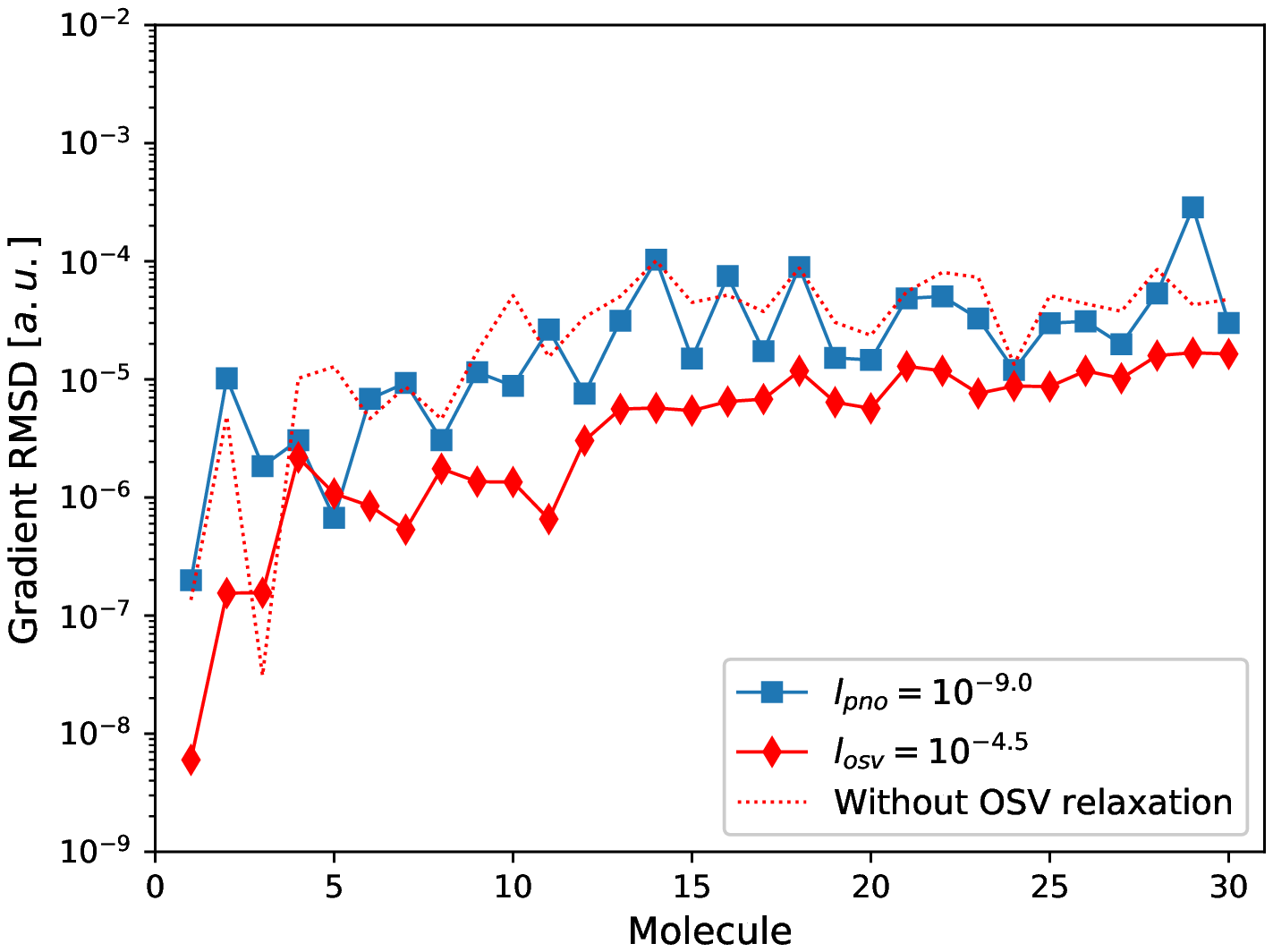}
  \caption*{(d)}
 \end{minipage}
 \caption{
 Comparisons of the gradient RMSDs from the RI-MP2 reference for 
 (a) OSV-MP2 (solid);  (b)-(c) DLPNO-MP2 (blue square) and OSV-MP2 (red diamond).
 THE OSV-MP2 gradients without OSV relaxation are presented in dotted lines.
 The molecules are taken from the Baker test set and ordered according to the number of atoms\cite{pinski2019analytical}.
 The DLPNO-MP2 gradient data are available in the reference\cite{pinski2019analytical}.
 All calculations were  performed with the basis set def2-TZVP.
 }
 \label{fig:gra_osv}
\end{figure}

To assess the convergence of OSV-MP2 gradients with respect to the OSV selection thresholds,
the RMSDs between the gradients of OSV-MP2 and RI-MP2 reference are presented 
in Figure~\ref{fig:gra_osv} for molecules of varying sizes and bonding types 
in the Baker test set\cite{baker1993techniques}.
As shown in Figure~\ref{fig:gra_osv}(a),
the average RMSDs among all computed molecules are $1.3\times10^{-4}$, $3.2\times10^{-5}$ and $6.3\times10^{-6}$
for $l_{osv}=10^{-3.5}$, $10^{-4}$ and $10^{-4.5}$, respectively.
For $l_{osv}=10^{-4.0}$, 
the RMSDs range from $10^{-5}$--$10^{-7}$ 
for smaller molecules (the molecule number lower than 15), and 
increase to about $5\times10^{-4}$--$5\times10^{-5}$ for larger molecules. 

The effect of the OSV relaxation is illustrated in Figs.~\ref{fig:gra_osv}(b)-(d)
by comparing the OSV-MP2 gradients computed with and without OSV relaxation. 
The OSV-MP2 analytical gradient without OSV relaxation merely considers the 
MO- and AO-specific gradient contributions described in Secs.~\ref{sec:mogra} and \ref{sec:aogra}.
It is obvious that the inclusion of the OSV relaxation considerably reduces the RMSDs 
by an order of magnitude. For instance, with $l_{osv}=10^{-4}$ (Figure~\ref{fig:gra_osv}(c)),
the average RMSDs decrease from around $10^{-4}$ to $10^{-5}$. 
Nevertheless, the exclusion of OSV relaxations appears less significant
when more OSVs are selected according to $l_{osv}=10^{-4.5}$ (Figure~\ref{fig:gra_osv}(d)) 
by which the resulting gradient RMSDs are less than $10^{-4}$, 
virtually comparable to results with $l_{osv}=10^{-4}$ (Figure~\ref{fig:gra_osv}(c)).

The gradient RMSDs of OSV-MP2 are compared with those of DLPNO-MP2 available in
a recent publication\cite{pinski2019analytical}.  To be as consistent as
possible with the corresponding PNO thresholds ($l_{pno}=10^{-7}$, $10^{-8}$ and
$10^{-9}$), we adopted the OSV threshold as $l_{osv}=\sqrt{l_{pno}}$ for
comparison since the PNOs are chosen according to eigenvalues of semi-canonical
pair density matrices, that is about the squared eigenvalues of the associated
semi-canonical amplitudes.  Nonetheless, a rigorous accuracy comparison between
DLPNO-MP2 and OSV-MP2 is difficult, since at the same level of truncation (e.g,
$l_{pno}=10^{-8}$ vs $l_{osv}=10^{-4}$) nondiagonal pair amplitudes are
represented in a much more compact basis in the DLPNO approach than in the OSV
approach.

As seen in Figure~\ref{fig:gra_osv}(b) by comparing loose OSVs and PNOs, the
RMSDs of two methods are generally similar especially for larger molecules, yet
with marginally better performance for OSV-MP2 than DLPNO-MP2 for smaller
molecules.  For $l_{osv}=10^{-4.5}$/$l_{pno}=10^{-9}$ in
Figure~\ref{fig:gra_osv}(d), the RMSDs of OSV-MP2 are remarkably smaller than
those of DLPNO-MP2.  We note that benzidine ( molecule 29) is peculiar here for
DLPNO-MP2 with an RMSD above $10^{-4}$ even using $l_{pno}=10^{-9}$.  The
OSV-MP2 analytical gradient however yields no significant RMSDs which are
consistently below $10^{-4}$ and $10^{-5}$ for $l_{osv}=10^{-4.0}$ and
$10^{-4.5}$.

\subsection{Optimized molecular structures}

\subsubsection*{Bond lengths}

The statistical errors of OSV-MP2 bond lengths relative to the reference data of RI-MP2  
are summarized in Table~\ref{t_osv_len}. 
For all basis sets, tighter OSV thresholds lead to decreased errors of  bond lengths.
Notably, the OSV-MP2 optimization with $l_{osv}=10^{-4}$ is sufficiently accurate 
and increasing basis set sizes only slightly increases MAEs.
However, the calculations with $l_{osv}=10^{-3}$ yields much larger errors.
\begin{table}
  \caption{Mean error (ME), mean absolute error (MAE) and the maximum error (max) in bond lengths (pm) of 
           selected Baker test molecules.}
  \label{t_osv_len}
  \begin{tabular}{llccc}
    \hline
    $l_{osv}$	&		&	def2-SVP	&	def2-TZVPP	&	def2-QZVPP	\\
    \hline
    \(10^{-3}\)	&	ME	&	0.071	&	0.097	&	0.151	\\
    	&	MAE	&	0.081	&	0.166	&	0.167	\\
    	&	max	&	0.380	&	0.570	&	0.620	\\
    \(10^{-4}\)	&	ME	&	0.009	&	0.013	&	0.016	\\
    	&	MAE	&	0.014	&	0.017	&	0.019	\\
    	&	max	&	0.050	&	0.070	&	0.080	\\
    \(10^{-5}\)	&	ME	&	0.000	&	-0.002	&	0.005	\\
    	&	MAE	&	0.009	&	0.016	&	0.012	\\
    	&	max	&	0.040	&	0.080	&	0.060	\\
    \hline
  \end{tabular}
\end{table}

\begin{figure}[H]
 \begin{minipage}{8cm}
  \centering
  \includegraphics[width=8cm]{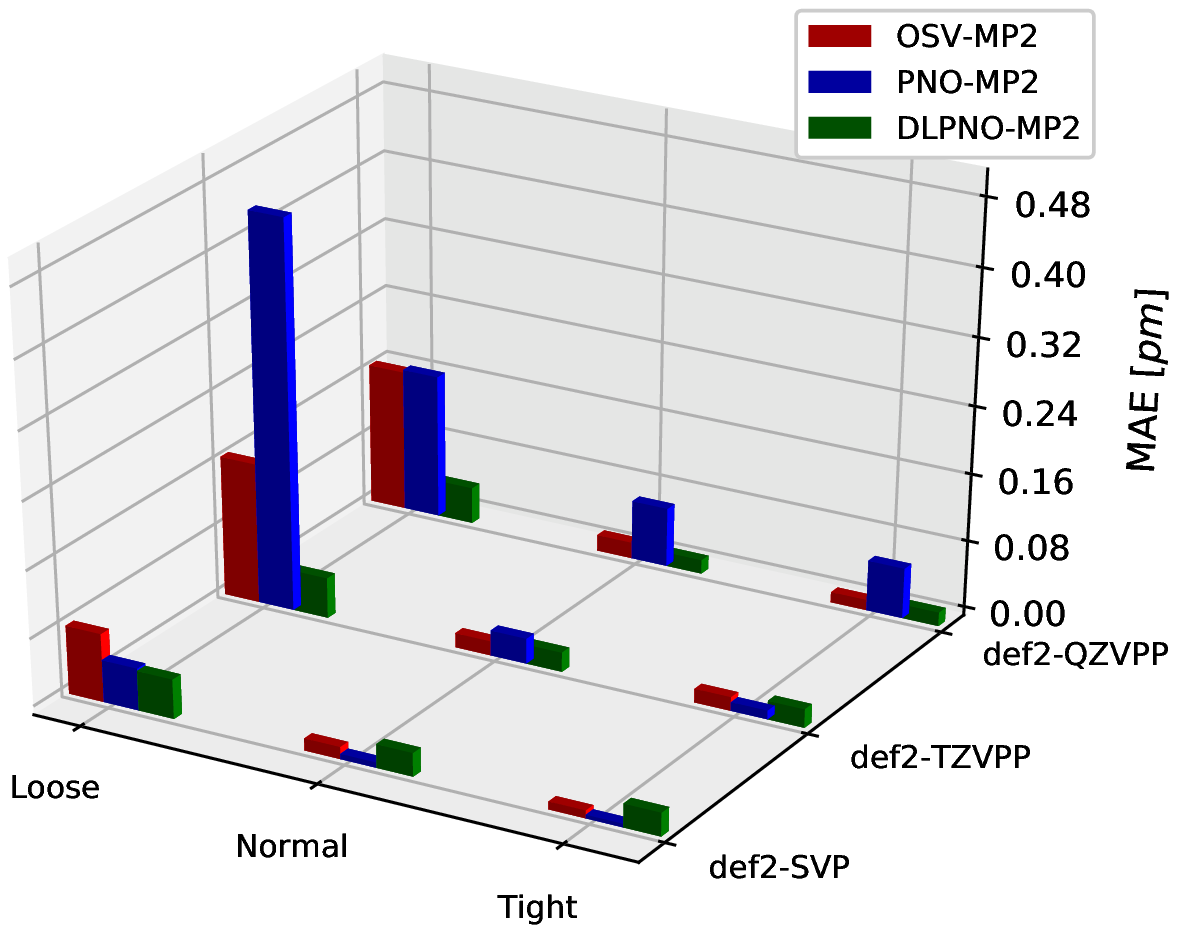} 
  \caption*{(a)}
 \end{minipage}
 \begin{minipage}{8cm}
  \centering
  \includegraphics[width=8cm]{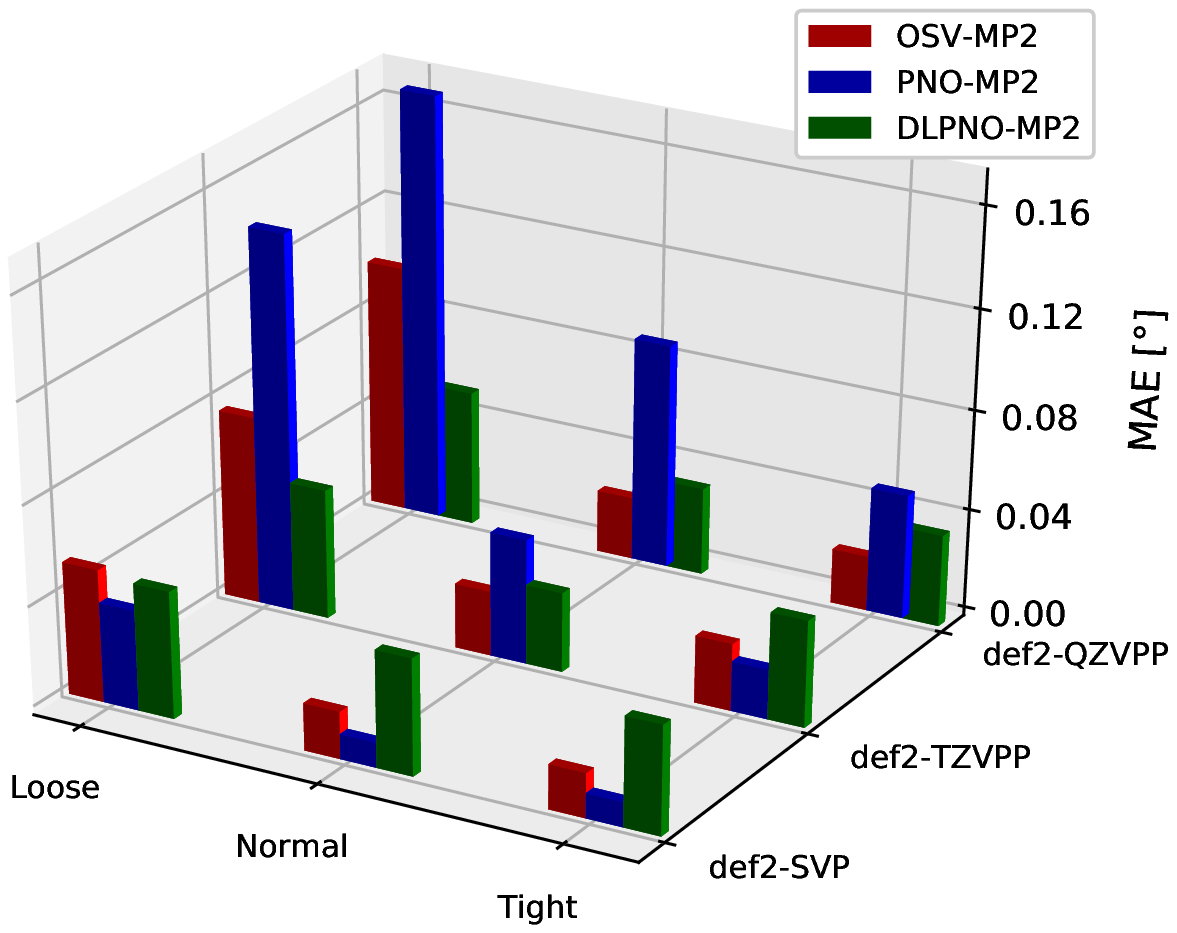}
  \caption*{(b)}
 \end{minipage}
  \caption{ Comparison of the MAEs in bond lengths (a) and angles (b) between
OSV-MP2, PNO-MP2 and DLPNO-MP2.  PNO-MP2 results in Ref.\cite{frank2017pno} were
computed without the relaxation of PNOs.  The loose ($l_{osv}=10^{-3.5}$ and
$l_{pno}=10^{-7}$), normal ($l_{osv}=10^{-4.0}$ and $l_{pno}=10^{-8}$) and tight
($l_{osv}=10^{-4.5}$ and $l_{pno}=10^{-9}$) selection thresholds are utilized.
The full PAO domains (TCutDO=0) were used for DLPNO-MP2 optimization with
Foster-Boys localization\cite{foster1960canonical} and RIJCOSX integrals.}
  \label{fig:com_len}
\end{figure}

In Figure~\ref{fig:com_len}(a),
the MAEs of bond lengths with different basis sets and OSV/PNO selection thresholds
are compared between OSV-MP2, PNO-MP2 and DLPNO-MP2.
DLPNO-MP2 yields lower MAEs than OSV-MP2 with the loose threshold for all basis sets,
but is overtaken by OSV-MP2 with tighter thresholds. 
For def2-TZVPP and loose threshold, the MAE for OSV-MP2 is significantly 
lower than PNO-MP2 by around 0.3 pm, but larger than DLPNO-MP2. 
The performances of the three methods are comparable for normal and tight calculations.

By repeating the OSV-MP2 geometry optimization, the long interatomic distances 
of noncovalent bonds  have been examined in Table~\ref{tab:bond_long}.
We have chosen DTFS and RESVAN molecules out of LB12 set for which 
the DLPNO-MP2  optimized Si-N and S-S bond distances
report quite large errors  in Ref.\cite{pinski2019analytical}.
All electrons are correlated in OSV-MP2 calculations, 
and both OSV-MP2 geometries are well converged for three OSV-MP2 thresholds.
It is observed that $l_{osv}=10^{-4.5}$ 
is necessary in order to reduce the errors below 1.0 pm.
However, $l_{osv}=10^{-4.0}$
appears to be sufficient for achieving relative deviations below $1\%$, 
which is acceptable for such long bond distances.
In general, the OSV-MP2 outperforms DLPNO-MP2 for loose selection, and both methods are comparable
for normal and tight selections. 

\begin{table}[H]
  \caption{Comparisons of the optimized interatomic distances in DTFS and RESVAN molecules between  OSV-MP2
           and DLPNO-MP2.  
           The long noncovalent interatomic distances are specified in Ref.\cite{pinski2019analytical}
}
  \label{tab:bond_long}
  \begin{tabular}{llcccc}
    \hline											
	&		&	\multicolumn{2}{c}{DTFS (Si-N)}			&	\multicolumn{2}{c}{RESVAN (S-S)}	\\\cline{3-4}\cline{5-6}
	&	Threshold$^a$	&	$\Delta$r (pm)	&	r\textsubscript{RI-MP2} (pm)	&	$\Delta$r (pm)	&	r\textsubscript{RI-MP2} (pm)	\\
    \hline											
DLPNO-MP2$^b$	&Loose	&	1.94	&	214.91$^b$	&	9.36	&	390.66$^b$	\\
		&Normal	&	0.78	&			&	2.98	&		\\
		&Tight	&	0.33	&			&	0.91	&		\\
OSV-MP2$^b$	&Loose	&	0.70	&	211.80$^c$	&	6.90	&	385.80$^c$	\\
		&Normal	&	0.40	&			&	2.70	&		\\
		&Tight	&	0.00	&			&	0.80	&		\\
    \hline
  \multicolumn{6}{l}{$^a$ Predefined in Figure~\ref{fig:com_len}.} \\
  \multicolumn{6}{l}{$^b$ DLPNO-MP2 results with frozen core approximation from Ref.\cite{pinski2019analytical}.} \\ 
  \multicolumn{6}{l}{$^c$ Our results without frozen core approximation.} 
  \end{tabular}
\end{table}

\subsubsection*{Bond and dihedral angles}

The errors of bond angles are reported in Table~\ref{tab:osv_ang} for selected Baker's test molecules 
according to the specification in Ref.\cite{frank2017pno}. 
In general, the MAEs are smaller than $0.1^\circ$  for $l_{osv}$ all values in combination with all basis sets.
$l_{osv}=10^{-3}$ results in relatively large maximum errors about $1.0^\circ$.
Both $l_{osv}=10^{-4}$ and $10^{-5}$ substantially reduce the maximum errors by about an order of magnitude
and are recommended for accurate structure optimizations.
The performances of OSV-MP2, PNO-MP2 and DLPNO-MP2 in bond angles are compared in Figure~\ref{fig:com_len}(b). 
Most notably PNO-MP2 without the PNO relaxation yields larger errors 
than OSV-MP2 and DLPNO-MP2, in particular for def2-TZVPP and def2-QZVPP basis sets.
The performances of OSV-MP2 and DLPNO-MP2 are similar with normal and tight thresholds. 

\begin{table}[H]
  \caption{Mean absolute error (MAE) and the maximum error (max) in bond angles ($^\circ$) of selected Baker test molecules.}
  \label{tab:osv_ang}
  \begin{tabular}{llccc}
    \hline
    $l$\textsubscript{osv}	&		&	def2-SVP	&	def2-TZVPP	&	def2-QZVPP	\\
    \hline
    \(10^{-3}\)	&	MAE	&	0.05 	&	0.08 	&	0.10 	\\
    	&	max	&	1.10 	&	0.80 	&	1.00 	\\
    \(10^{-4}\)	&	MAE	&	0.02 	&	0.03 	&	0.03 	\\
    	&	max	&	0.10 	&	0.10 	&	0.20 	\\
    \(10^{-5}\)	&	MAE	&	0.02 	&	0.03 	&	0.02 	\\
    	&	max	&	0.10 	&	0.30 	&	0.20 	\\
    \hline
  \end{tabular}
\end{table}

The dihedral angles of benzidine molecule are compared between OSV-MP2, PNO-MP2 and DLPNO-MP2 in Table~\ref{tab:com_di}. 
Overall, OSV-MP2 performs much better than PNO-MP2 and DLPNO-MP2 for all thresholds and basis sets,
and the deviations from RI-MP2 dihedral angles are less than $0.2^\circ$ for OSV-MP2/normal and OSV-MP2/tight.

\begin{table}[H]
  \caption{Dihedral angles of benzidine for OSV-MP2, PNO-MP2 and DLPNO-MP2 with the basis sets of  def2-TZVPP and def2-QZVPP.}
  \label{tab:com_di}
  \begin{tabular}{llcccc}
    \hline
    Basis set	&Threshold\(^a\)&	RI-MP2	&	OSV-MP2	&PNO-MP2\(^b\)	&	DLPNO-MP2	\\
    \hline
    def2-SVP	&	Loose 	&	138.8	&	137.2	&-		&	-\(^c\)	\\
        	&	Normal	&		&	138.7	&-		&	137.5	\\
        	&	Tight	&		&	138.7	&-		&	137.6	\\
    def2-TZVPP	&	Loose	&	142.2	&	139.9	&92.6		&	-\(^c\)	\\
        	&	Normal	&		&	142.0	&143.2		&	138.9	\\
        	&	Tight	&		&	142.2	&142.3		&	139.3	\\
    def2-QZVPP	&	Loose	&	142.1	&	138.7	&138.4		&	-\(^c\)	\\
    	        &	Normal	&		&	141.9	&141.3		&	139.2	\\
    	        &	Tight	&		&	142.1	&140.6		&	139.5	\\
    \hline
  \multicolumn{6}{l}{$^a$ Predefined in Figure~\ref{fig:com_len}.} \\
  \multicolumn{6}{l}{$^b$ PNO-MP2 results without PNOs relaxation from Ref.\cite{frank2017pno}.} \\ 
  \multicolumn{6}{l}{$^c$ DLPNO-MP2 reported not converged.} 
  \end{tabular}
\end{table}

\subsubsection*{Performance with pair screening}

The use of pair screening can considerably accelerate the OSV-MP2 calculations by discarding 
the pairs of occupied orbitals that make little contribution to the total correlation energy. 
By exploring the orbital locality and the definition of OSVs,
the OSV overlap matrix elements associated with a pair $(i,j)$ exhibit an exponential decay 
with the separation between $i$ and $j$.
Therefore the relevant pairs entering OSV-MP2 calculations 
are chosen according to the previous simple scheme\cite{yang2011tensor} 
in which the renormalized OSV overlap matrix is computed for a given $(i,j)$ pair and compared to 
a predefined pair screening threshold $l_{pair}$.
When a looser $l_{pair}$ (greater value) is used, more orbital pairs 
will be screened and not participate in the OSV-MP2 energy and gradient computation. 
The MAEs of bond lengths, bond angles and dihedral angles are reported 
with respect to $l_{pair}$ in Table~\ref{tab:scr}. 
It is shown that the MAEs at \(l_{pair}=10^{-4}\) are similar to those without pair screening
for both bond lengths and angles. 
However, there is a significant increase of MAEs as $l_{pair}$ is increased
from \(10^{-4}\) to \(10^{-2}\).
Interestingly, for dihedral angles, the errors 
for all $l_{pair}$ thresholds are less than $0.8^\circ$. 
 
\begin{table}[H]
  \caption{Mean absolute error (MAE) of bond lengths, bond angles and absolute error (AE) of dihedral angles 
with respect to pair screening thresholds.}
  \label{tab:scr}
  \begin{tabular}{lccc}
    \hline							
	&	Bond length	&	Bond angle	&	Dihedral angle	\\\cline{2-4}
							
    $l_{pair}$	&	MAE (pm)	&	MAE ($^{\circ}$)	&	AE ($^{\circ}$)	\\
    \hline							
    $10^{-2}$	&	0.043	&	0.043	&	0.6	\\
    $10^{-3}$	&	0.021	&	0.028	&	0.3	\\
    $10^{-4}$	&	0.019	&	0.023	&	0.5	\\
	    0	&	0.019	&	0.025	&	0.2	\\
    \hline							
  \end{tabular}
\end{table}

\subsubsection*{Timing comparison}
We finally compare the elapsed times between RI-MP2, DLPNO-MP2 and  OSV-MP2 for both energy
and gradient evaluations on a single CPU.  For all molecules considered in
Table~\ref{tab:timing}, our current OSV-MP2 implementation achieves speedups of
3-10 folds for gradients and 0.4-4.0 for energies compared to RI-MP2, respectively. In particular,
the OSV-MP2 gradient computation is faster than RI-MP2 by an order of magnitude
for the longest molecule (Gly)$_{14}$. For Nonactin molecule similar to
(Gly)$_{14}$ in size, OSV-MP2 gradient calculation exhibits a poorer speedup
than (Gly)$_{14}$ due to more kept pairs of Nonactin (6737 out of 20100 pairs)
than (Gly)$_{14}$ (4218 out 23220 pairs), since apparently  the pair screening
is less effective to the cyclic Nonactin structure than the linear (Gly)$_{14}$.
The average pair domain sizes of Nonactin and (Gly)$_{14}$ are similar, i.e.,
both own 96 OSVs, which is much larger than DLPNO pair domains (about 17-20 PNOs).
On the other hand, DLPNO-MP2 retains 14937 and 8192 pairs for (Gly)$_{14}$ and
Nonactin, respectively, that are much larger than those of OSV-MP2.
Moreover, the OSV-MP2 energy and gradient scalings are $N^{2.74}$ and
$N^{2.96}$, respectively, as shown in Figure~S1. This is still higher than
DLPNO-MP2 and leads to longer elapsed time than DLPNO-MP2 by nearly 2 folds for
large glycine chains.  However, for smaller molecules, OSV-MP2 gradient
computation can be 2 times faster than DLPNO-MP2, which makes it attractive for
driving efficient BOMD simulations on molecules of similar size.
\begin{table}[H]
  \caption{
Timing comparisons for polyglycine chains (Gly)$_n$ and Nonactin
(C$_{40}$H$_{64}$O$_{12}$) with respect to the number of basis ($N$) and
auxiliary functions ($N_{aux}$) between RI-MP2 and OSV-MP2 computations. Elapsed
times (minutes) of a single-point energy and gradient are reported for each
molecule in the first and second row entry, respectively. Percentages of
correlation energy recovery are reported for DLPNO-MP2 and OSV-MP2 as the first
and second element in the last Column, respectively.  These calculations were
carried out serially on a single CPU (Intel Xeon E5-2640 v3@2.60GHz) with def2-TZVP
basis.  $l_{osv}=10^{-4}$ and $l_{pair}=10^{-3}$ were used for OSV selection and
pair screening. RI-MP2 was performed on the quantum chemistry program
ORCA\cite{neese2018software}. For DLPNO-MP2/RIJCOSX, TCutPNO=$10^{-8}$ and
TCutDO=$10^{-2}$ were set for constructing PNOs and associated PAO domains with
Foster-Boys localization\cite{foster1960canonical}, respectively.}
\label{tab:timing}
\begin{tabular}{cccccccc}
\hline
Molecules	& $N$	& $N_{aux}$ & RI-MP2       & DLPNO-MP2       & OSV-MP2       & Speedups & Percentages
\\
\hline														
(Gly)$_4$	&611	&1502	&2     & 7    &6     & 0.4  & 99.96\%, 99.96\% \\
                &       &       &49    & 28   &14    & 3.5  &         \\[1mm]
(Gly)$_6$	&895	&2200	&14    & 14   &15    & 0.9  & 99.96\%, 99.95\% \\
                &       &       &163   & 65   &40    & 4.1  &         \\[1mm]
(Gly)$_8$	&1179	&2898	&53    & 22   &32    & 1.7  & 99.96\%, 99.95\% \\
                &       &       &475   & 112  &90    & 5.3  &         \\[1mm]
(Gly)$_{10}$	&1463	&3596	&145   & 30   &65    & 2.2  & 99.95\%, 99.95\% \\
                &       &       &1069  & 171  &198   & 5.4  &         \\[1mm]
(Gly)$_{12}$	&1747	&4294	&355   & 39   &112   & 3.2  & 99.95\%, 99.95\% \\
                &       &       &2796  & 243  &332   & 8.4  &         \\[1mm]
(Gly)$_{14}$	&2031	&4992	&751   & 52   &194   & 3.9  & 99.95\%, 99.95\% \\
                &       &       &5937  & 328  &569   & 10.4 &         \\[1mm]
Nonactin  	&1996	&4912	&598   & 135  &198   & 3.0  & 99.91\%, 99.89\% \\
                &       &       &4728  & 598  &697   & 6.8  &         \\
\hline
  \end{tabular}
\end{table}

\section{OSV-MP2-DRIVEN \textit{AB-INITIO} BOMD}

\subsection{Protonated Eigen and Zundel water cations}

We have performed the constant $NVE$ simulation for protonated Eigen  
(H\textsubscript{9}O\textsubscript{4}\textsuperscript{+}) and  
Zundel water cluster (H\textsubscript{13}O\textsubscript{6}\textsuperscript{+}).
They are not only structural units of biological and chemical significance,
but also the benchmark systems that have been extensively used to establish accuracy of other theories.

\subsubsection*{Energy drifts}

In OSV-MP2 $NVE$ simulation, 
the OSV-MP2 approximated trajectories propagate according to the numerical integration over a finite time step
which may break the energy conservation by a range of drifts at long simulation time.
Therefore such drifts must be examined carefully with respect to both OSV and pair selections.
The results are reported in Table~\ref{tab:enedrift} for benchmarking OSV-MP2 BOMD accuracy.
When no pairs are screened ($l_{pair}=0.0$), all energy drifts are very small.
The total energies of all OSV-MP2/10 ps trajectories with $l_{osv}=10^{-3}$ 
are conserved within 1.0 kJ/mol,
the energy drifts are substantially reduced with $l_{osv}=10^{-4}$
by two and one orders of magnitude for 
H\textsubscript{9}O\textsubscript{4}\textsuperscript{+}
and H\textsubscript{13}O\textsubscript{6}\textsuperscript{+}, respectively. 
The RMSDs, which measure the time-dependent energy fluctuation statistically,
are as small as half kJ/mol for $l_{osv}=10^{-3}$ and $0.1$--$0.2$ kJ/mol for $l_{osv}=10^{-4}$.
The difference of the computed $T_{\mathrm{av}}$ between $l_{osv}=10^{-3}$ and $l_{osv}=10^{-4}$
is about 1 K for  H\textsubscript{9}O\textsubscript{4}\textsuperscript{+} and
5 K for  H\textsubscript{13}O\textsubscript{6}\textsuperscript{+}, respectively.

Table~\ref{tab:enedrift} suggests that the use of pair screenings yields larger statistical errors than the OSV selection.
Nevertheless, a proper combination of selected  $l_{osv}$ and $l_{pair}$  
can produce results of acceptable accuracy. 
For instance, for $l_{osv}=10^{-4}$, the choice of the medium pair screening $l_{pair}=0.001$ 
does not lead to significant shifts of energy (both $\delta E$ and RMSD) and temperature. 
However, with $l_{pair}=0.01$ and $l_{pair}=0.02$,
the energy conservation is not well sustained.
As seen in Figure~S2,  with more pair screenings for Zundel cluster,
the $l_{osv}=10^{-4}/l_{pair}=0.01$ simulation after about 4.5 ps leads to a hotter Zundel cation
by 1 kJ/mol, probably arising from a more drastic change of the number of the kept pairs with time.

\begin{table}
  \caption{Comparison of statistical energy conservation properties with respect to OSV and pair selections 
           for $NVE$/6-31+g(d,p) simulation. 
           $T_{\mathrm{av}}$ is the average temperature computed according to the equipartition theorem 
           for the average kinetic energy.
           $\delta E $ is the energy drift that is the difference of the linear least-square fit 
           to all energies at the first and last time step.
           RMSDs are given among all energies relative to this linear fit.
          }
  \label{tab:enedrift}
  \begin{tabular}{c|ccccc}
    \hline
    Molecule & $l_{osv}$& $l_{pair}$  & $T_{\mathrm{av}}$ (K) & $\delta E$ (kJ/mol)& RMSD (kJ/mol)\\
    \hline
    H$_{9}$O$_{4}^{+}$ & $10^{-3}$   & 0.000  &149.3 & -0.45  & 0.41     \\
    &$10^{-3}$ & 0.001   &     149.9  &  -0.50 & 0.40 \\
    &$10^{-3}$ & 0.010&   152.0 &  0.94 &  0.65   \\
    &$10^{-3}$  & 0.020 &152.1 &  1.21  &1.21  \\

    &$5.0\times 10^{-4}$  & 0.000 &   149.3 &  0.36&  0.32 \\
 &$10^{-4}$   & 0.000 & 150.4 &0.00 & 0.17 \\
    &$10^{-4}$  & 0.001 & 149.5 & -0.04  &0.18 \\
   &$10^{-4}$ & 0.010 &150.8  &  0.08 &0.13 \\
   &$10^{-4}$  & 0.020&  149.1& 0.08 & 0.14  \\
    \hline
    H$_{13}$O$_{6}^{+}$ & $10^{-3}$ & 0.000& 153.0 & -0.98   &0.55 \\
     &  $10^{-3}$ & 0.001&191.1 & 50.97 &16.14   \\
     & $5.0\times 10^{-4}$  & 0.000 & 146.6  & -0.02  &0.21 \\
    & $10^{-4}$  &0.000 & 148.5  & 0.06  &0.22  \\
     & $10^{-4}$  &0.001 & 149.2  & -0.04  &0.25  \\
      & $10^{-4}$  &0.010 & 150.3  & 1.57 & 0.62 \\
        & $10^{-4}$  &0.020 & 150.0  & -0.53 & 0.34 \\
    \hline
  \end{tabular}
\end{table}

\subsubsection*{Radial distribution function (RDF)}

\begin{figure}
\begin{minipage}[t]{0.49\textwidth}
\centering
\includegraphics[width=\linewidth]{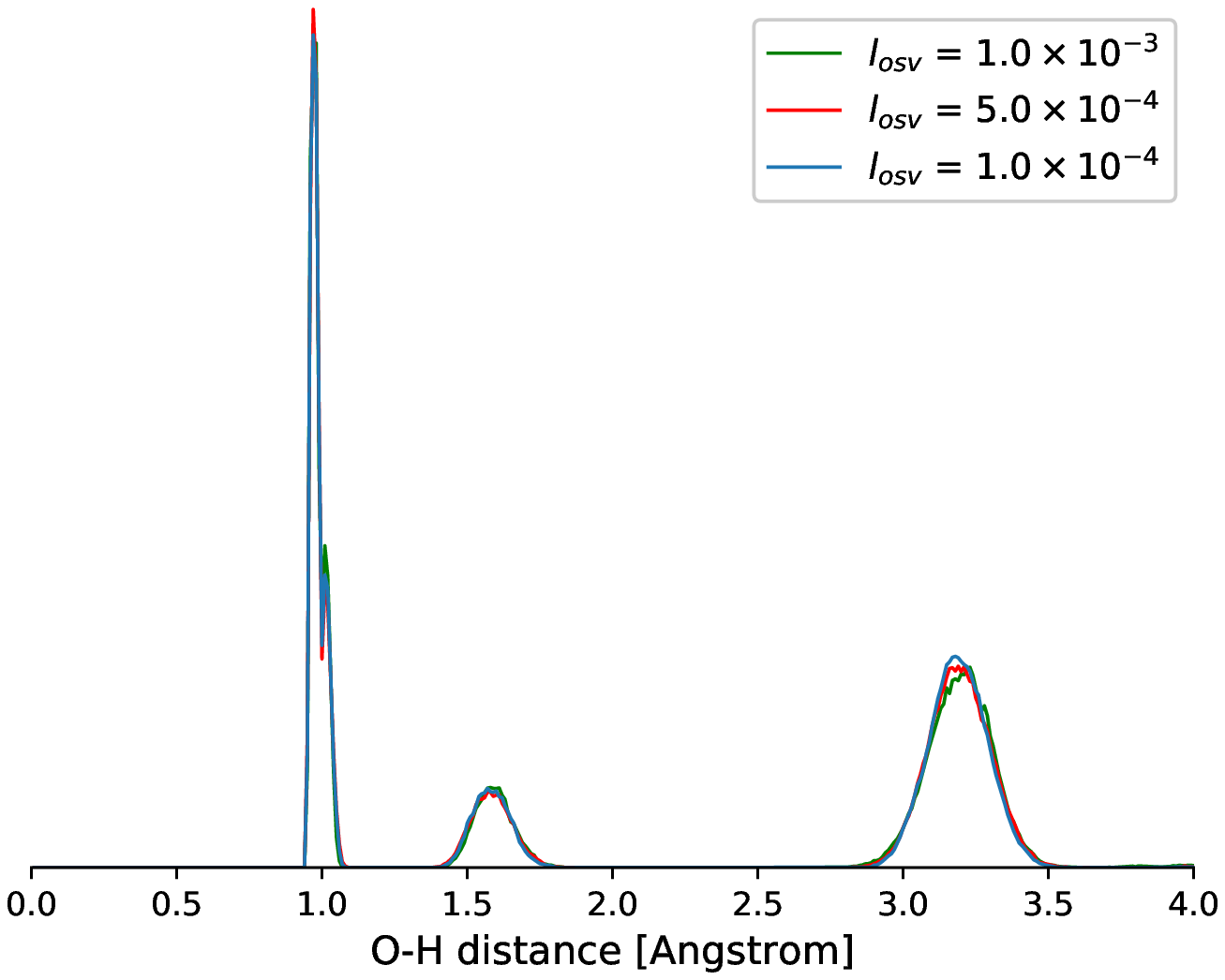}
  \caption*{(a) Eigen H\textsubscript{9}O\textsubscript{4}\textsuperscript{+}}
\end{minipage}
\begin{minipage}[t]{0.49\textwidth}
\centering
\includegraphics[width=\linewidth]{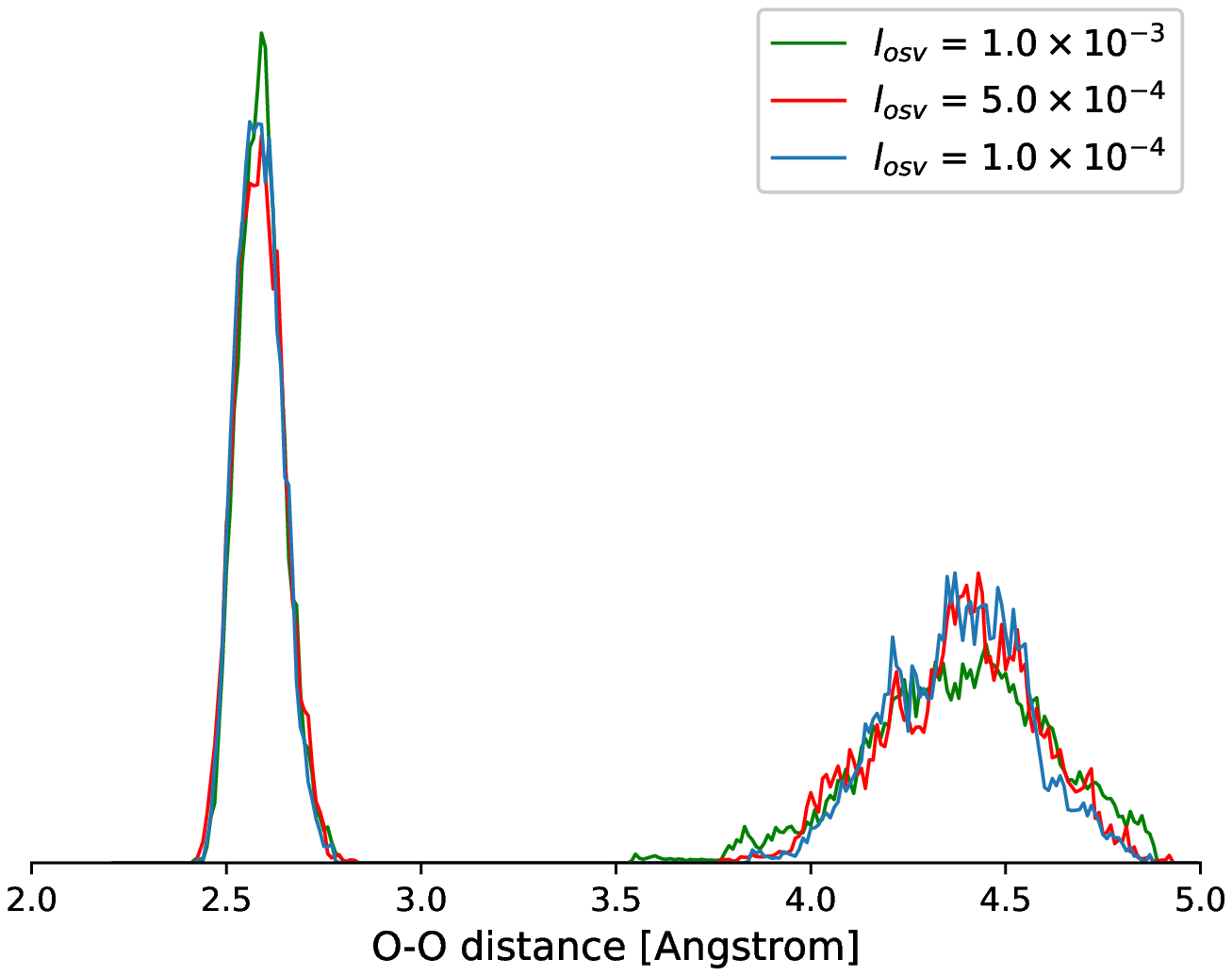}
  \caption*{(b) Eigen H\textsubscript{9}O\textsubscript{4}\textsuperscript{+}}
\end{minipage}
\begin{minipage}[t]{0.49\textwidth}
\centering
\includegraphics[width=\linewidth]{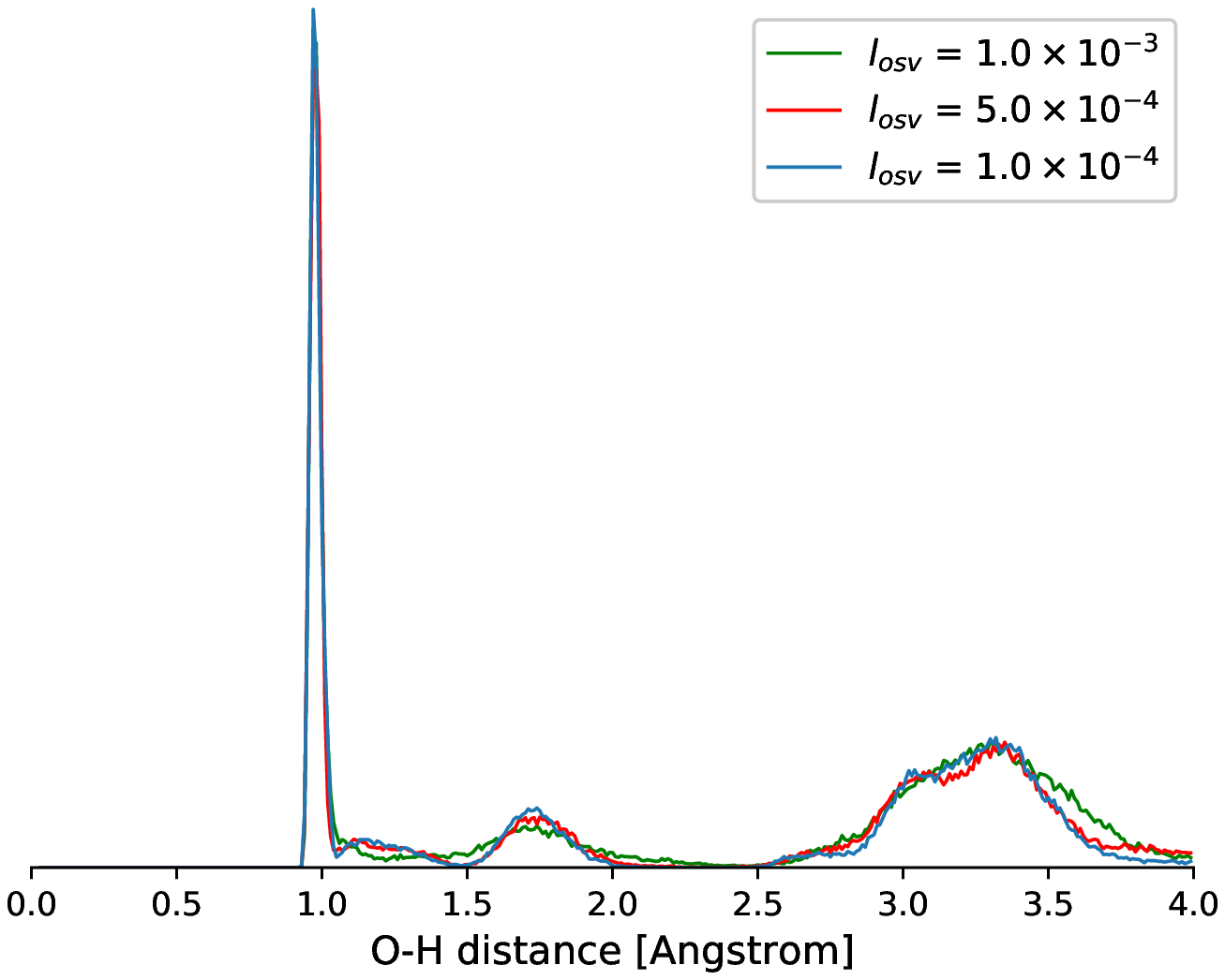}
  \caption*{(c) Zundel H\textsubscript{13}O\textsubscript{6}\textsuperscript{+}}
\end{minipage}
\begin{minipage}[t]{0.49\textwidth}
\centering
\includegraphics[width=\linewidth]{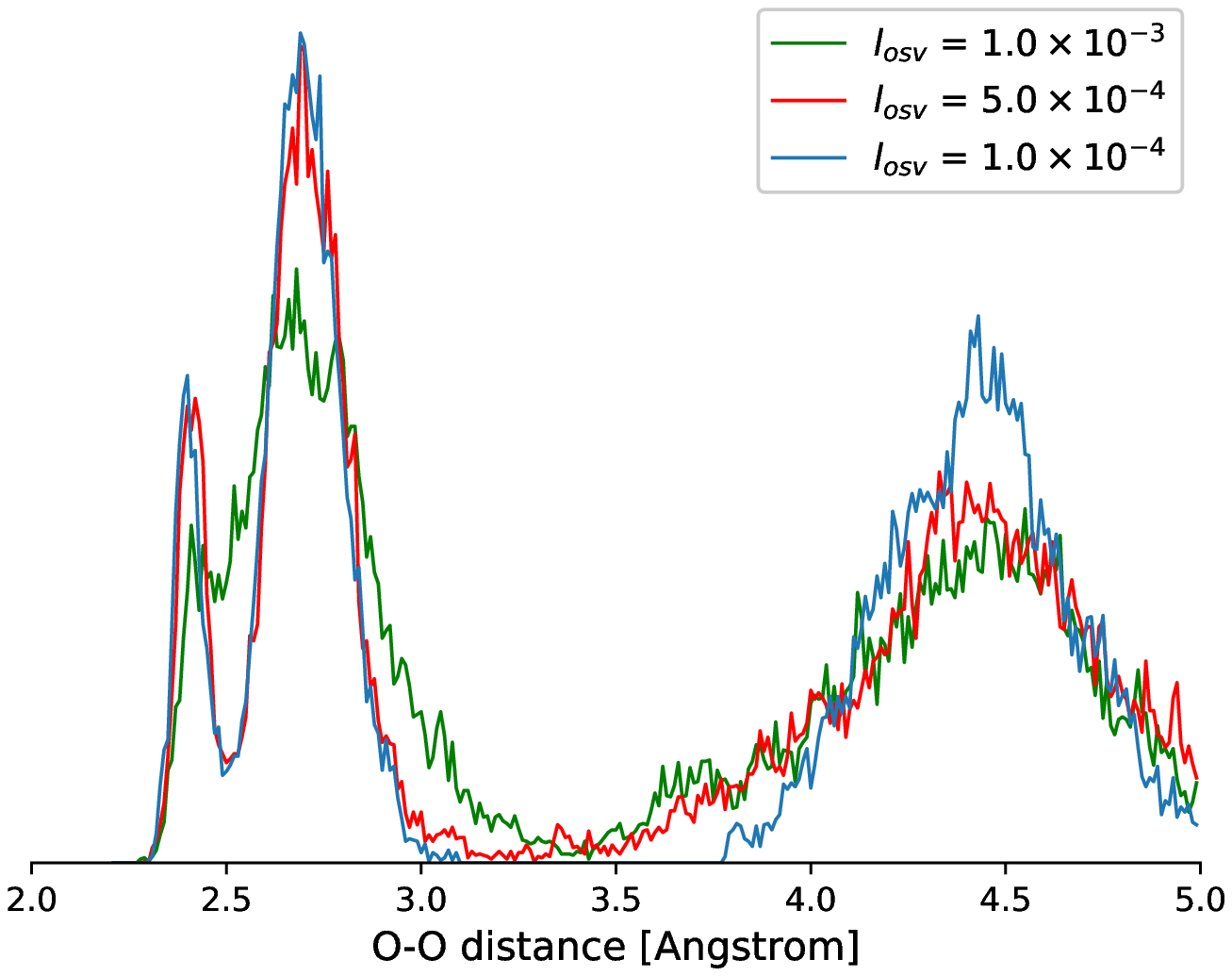}
  \caption*{(d) Zundel H\textsubscript{13}O\textsubscript{6}\textsuperscript{+}}
\end{minipage}
\caption{Radial distribution functions for O-H (left) and O-O (right) distances
         with respect to the OSV selection ($l_{osv}$) in the absence of pair screening 
         for (a-b) Eigen and (c-d) Zundel clusters.
         }\label{fig:rdflosv}
\end{figure}

The trajectory specification of computing RDFs of the O-O and O-H 
distances was adopted according to the description of 
Ref.\cite{li2016hybrid}. 
As seen in Figure~\ref{fig:rdflosv},
the OSV-MP2 BOMD calculations with $l_{osv}=10^{-4}$ are capable of 
retrieving all O-O and O-H structural details 
including  the RDF landscape and peak positions 
for both Eigen and Zundel clusters,
and also in excellent agreement with the canonical MP2 BOMD reference 
results\cite{li2016hybrid}. 
However,  for Zundel cluster, 
the calculations with the loose OSV selection $l_{osv}=10^{-3}$ 
do not well resolve two innermost peaks of the O-O RDF
at about $2.4$ \AA~(Zundel-like O-O distance) 
and $2.8$ \AA~(Eigen-like O-O distance),
but rather predict a more dominating Eigen-like solvation shell.
It is demonstrated in Figure~S3
that the pair screenings, when combined with the normal OSV 
selection $l_{osv}=10^{-4}$, have little effects on 
the RDF landscapes yet with a small broadening 
of the RDF peaks at longer O-H and O-O distances 
by increasing $l_{pair}$.

\subsubsection*{Vibrational density of states (VDOS)}

\begin{figure}
\begin{minipage}[t]{0.49\textwidth}
\centering
\includegraphics[width=\linewidth]{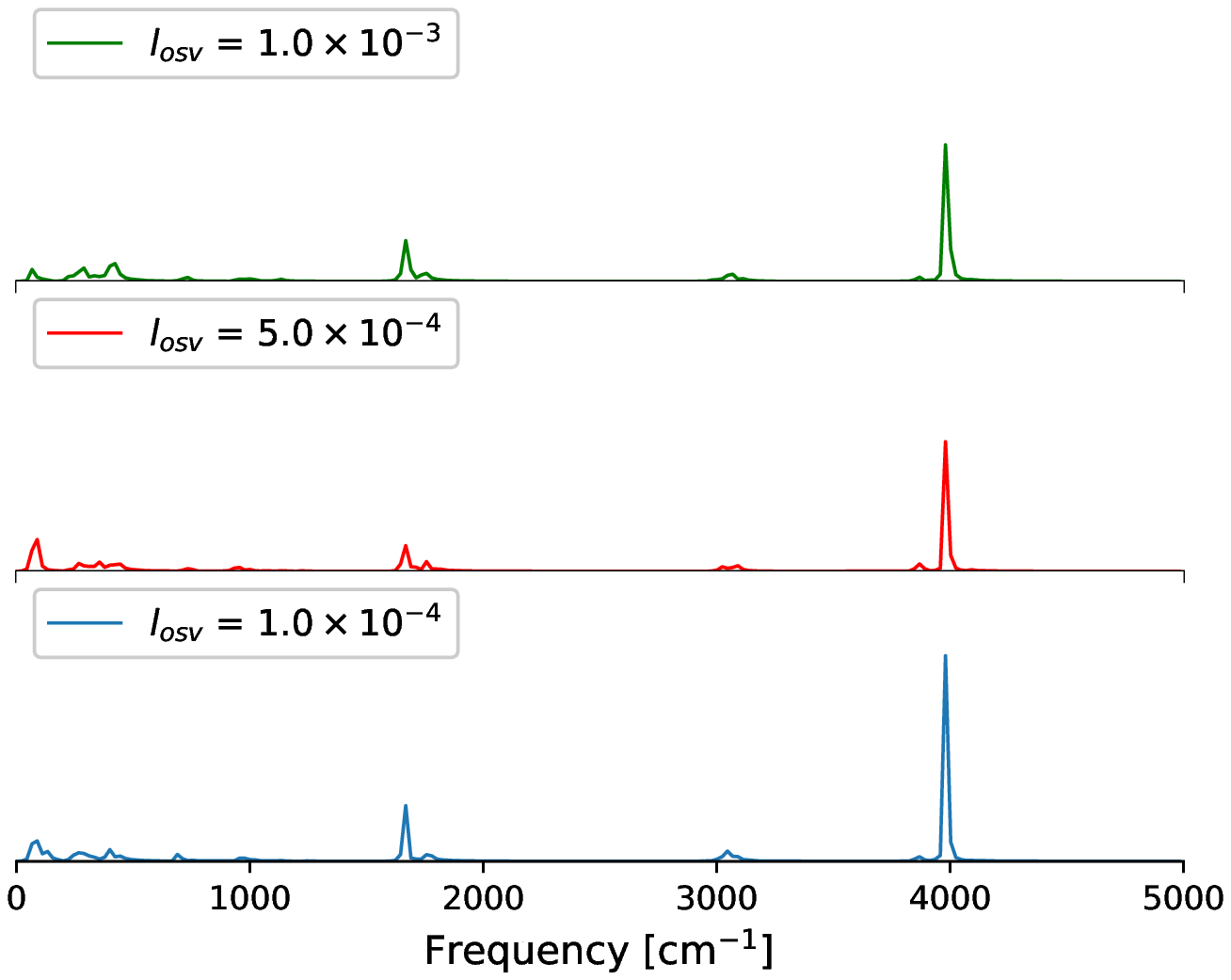}
  \caption*{(a) Eigen H\textsubscript{9}O\textsubscript{4}\textsuperscript{+}}
\end{minipage}
\begin{minipage}[t]{0.49\textwidth}
\centering
\includegraphics[width=\linewidth]{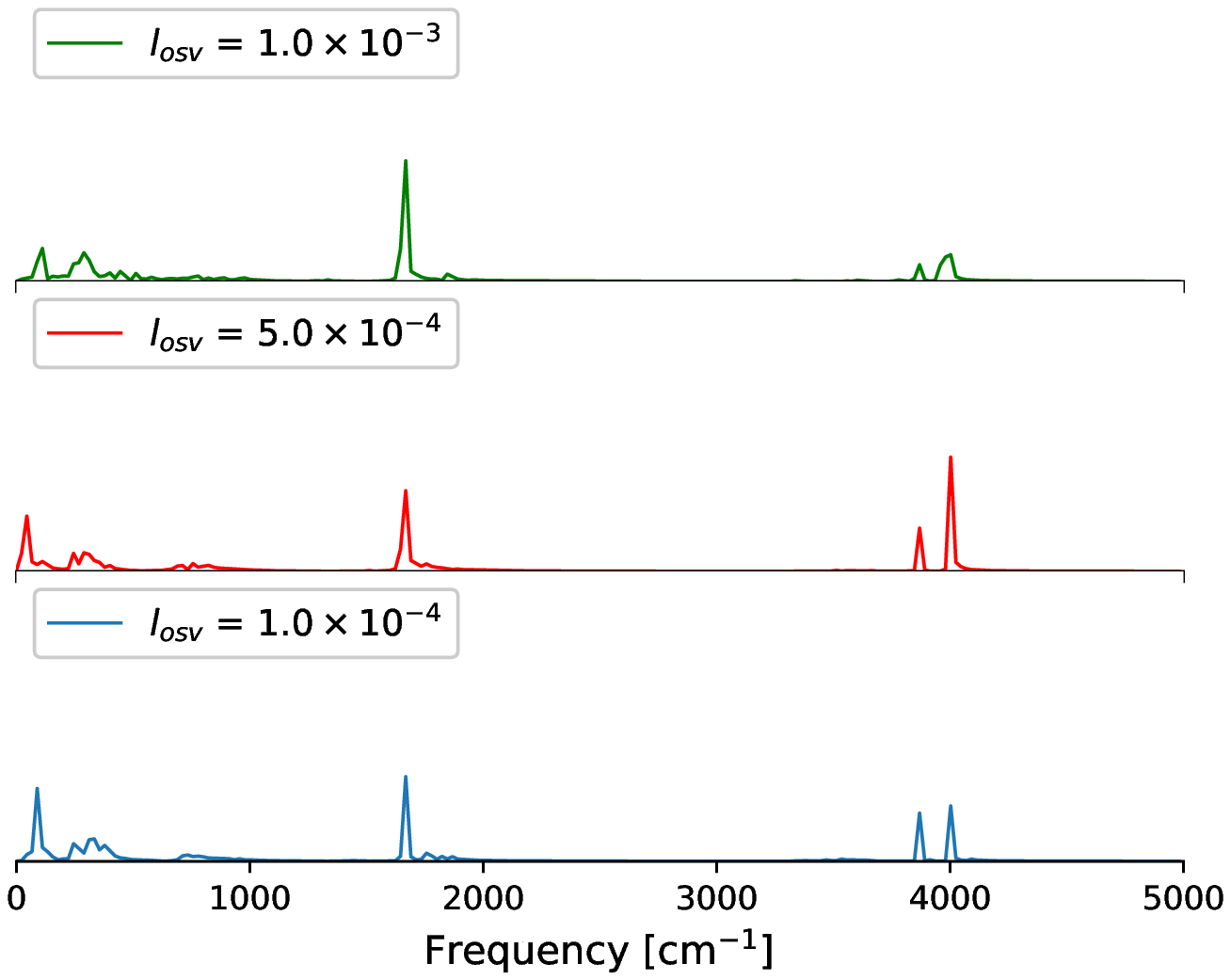}
  \caption*{(b) Zundel H\textsubscript{13}O\textsubscript{6}\textsuperscript{+}}
\end{minipage}
\caption{Vibrational density of states  with respect to the OSV selection $l_{osv}$
         for (a) Eigen and (b) Zundel clusters.
         }\label{fig:vdoslosv}
\end{figure}

Vibrational density of states are computed as the Fourier transform
of the velocity autocorrelation function
according to the Ref.\cite{li2016hybrid}. 
However, our initial structures are generated from RI-MP2 optimization,
with the momentum corresponding to 300 K.
We compare the computed VDOS spectra in Figures~\ref{fig:vdoslosv} and S4.
The positions of significant peaks can be hardly affected by the OSV 
selection and pair screening.
In particular, for Eigen cluster in Figures~\ref{fig:vdoslosv} (a) and S4(a), 
the weak peaks at about 3000 cm$^{-1}$ representing the proton stretch mode are well 
reproduced\cite{haycraft2017efficient} in all OSV-MP2 BOMD calculations.
For Zundel cluster in Figures~\ref{fig:vdoslosv} (b) and S4(b),
the two peaks of medium intensity around 4000 cm$^{-1}$ 
are clearly resolved.
However, the peak intensities are largely influenced 
by the combined $l_{osv}$ and $l_{pair}$. For instance,
the peaks at both low and high frequency regions are relatively 
intensified by decreasing $l_{osv}$.
On the other hand, a large pair screening appears to substantially
weaken the 4000 cm$^{-1}$ peak at the lower frequency side.

\subsection{Rotational free energy of ethanol}

The OSV-MP2/cc-pvTZ $NVT$ simulations were carried out for computing
the rotational free energies of the coupled hydroxyl and methyl groups
in ethanol molecule at 300 K. 
The $NVT$ simulation features thermal energy exchange which 
may compensate the electronic energy loss due to selected OSVs  
through adding a thermostat into Hamiltonian for coupling the system and reservoir. 
This thus opens up the feasibility of making OSV-MP2 BOMD available for simulating systems at a finite temperature. 
However, the detailed investigation on the interplay between the thermal coupling and the OSV selection is 
not the subject of this work and will be probed in future applications.
In the current work, the Nos\'e-Hoover thermostat was employed 
with the temperature coupling time constant of 100 fs. 
The simulation temperature is conserved within a drift of only -0.051 K
for $l_{osv}=10^{-4}$ and $l_{pair}=0.0$.

\begin{figure}
\begin{minipage}[t]{0.6\textwidth}
\centering
 \includegraphics[width=\textwidth]{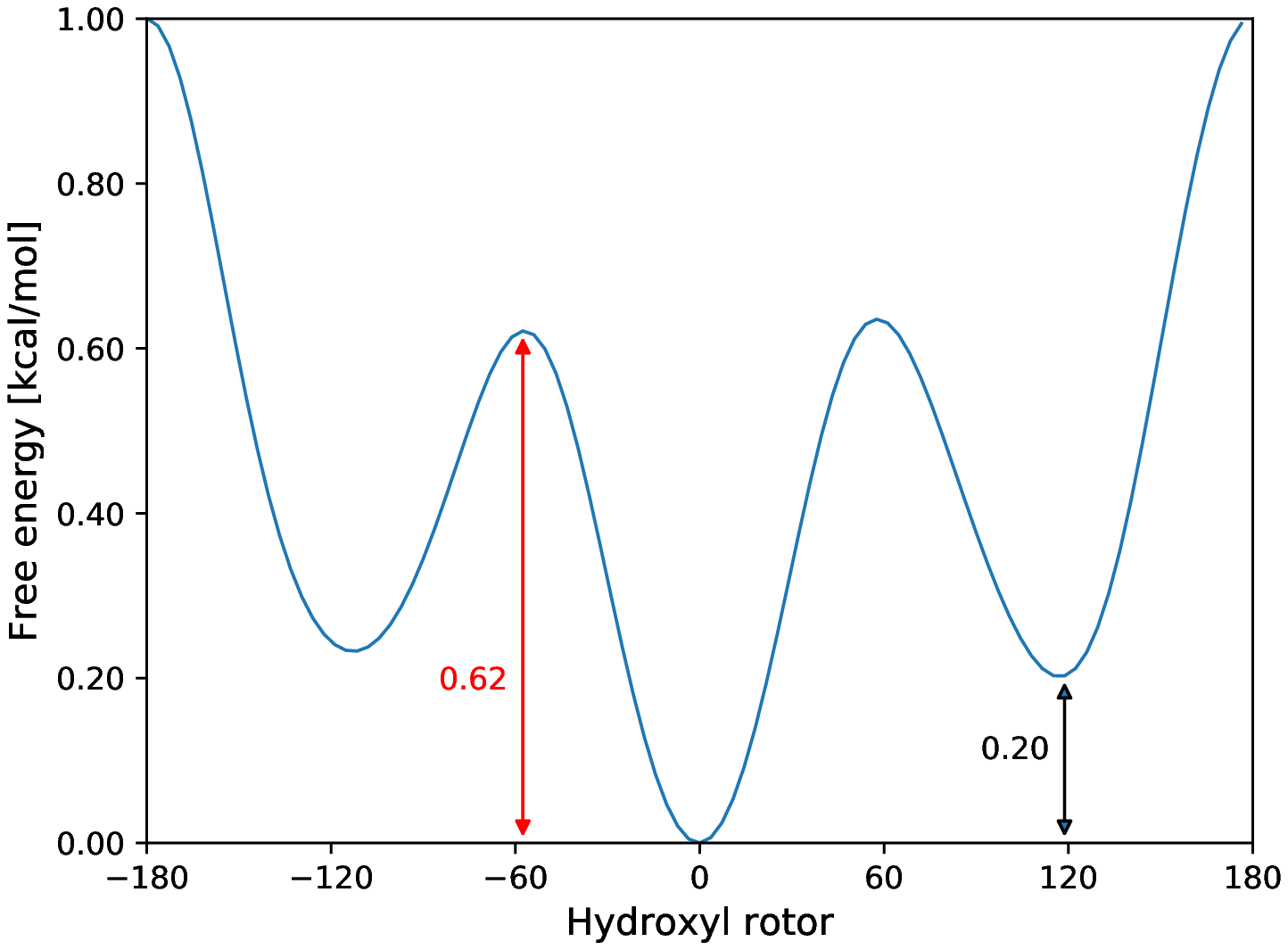}
\end{minipage}
 \caption{Free energy surface of ethanol for hydroxyl rotation}
 \label{fig:fes}
\end{figure}

Ethanol can exist in two conformers,
the trans-ethanol with the hydroxyl group trans to the methyl group,
and the  gauche-ethanol with the hydroxyl group gauche to the methyl group.
The gauche-ethanol stability computed by single point DFT is close to the trans-ethanol
by, for instances, 0.01 kcal/mol for B3LYP/cc-pVTZ\cite{durig2011r0} and -0.08 kcal/mol PBE-TS/cc-pVTZ\cite{chmiela2018towards}.
Our OSV-MP2/cc-pVTZ  $NVT$ simulation predicts that the trans-ethanol conformer is more stable than
the gauche-ethanol conformer by 0.22 kcal/mol, as shown in Figure~\ref{fig:fes}.
The OSV-MP2/cc-pVTZ $NVT$ also finds the free energy barriers of 1.00 kcal/mol and 0.62 kcal/mol 
to the hydroxyl rotation and trans-to-gauge transformation, respectively.
Recently, Chmiela et al.\cite{chmiela2018towards, sauceda2019molecular} reported that the corresponding CCSD(T) barriers are 
0.11 kcal/mol, 1.30 kcal/mol and 1.18 kcal/mol,
by training the symmetrized gradient-domain machine learning (sGDML) model
for the CCSD(T) force field in MD simulations.
It was observed however that the gauche is more stable than the trans 
by repeating the same calculation with sGDML@DFT(PBE-TS)\cite{chmiela2018towards}. 
Compared to sGDML@CCSD(T), our OSV-MP2 simulation seems to underestimate the energy level of the
transition state for tran-to-gauche transformation by 0.56 kcal/mol.
This disagreement may be ascribed to the difference of the levels in describing electron correlations 
between MP2 and CCSD(T) methods.
Nevertheless, single point  calculations\cite{dyczmons2004dimers} corresponding to 0 K 
show that the energy barriers are in fact similar between MP2 and CCSD(T), 
with differences of only a few hundredth kcal/mol.
Therefore
it remains a question whether such a  subtle difference of electron correlation between MP2 and CCSD(T) 
for ethanol has any significance due to a thermal fluctuation of  $\approx 0.6$ kcal/mol at 300 K. 
More importantly, we realize that in our computational setting for metadynamics simulation,
a relatively large time constant of 100 fs was used in order to achieve a small temperature drift ($\approx-0.051$ K)
and avoid poor coupling in a long time equilibration.
However, this inevitably results in a more wild distribution of Nos\'e-Hoover frequencies
and thus a larger thermal fluctuation.
More detailed studies on this issue within the OSV-MP2 framework are underway.

\section{CONCLUSIONS}

In this work, we have described the algorithm and implementation for analytically computing the 
energy derivatives from  all OSV-MP2 energy contributions with local molecular orbitals. 
We have shown that it is possible to evaluate the OSV relaxation 
by explicitly solving non-degenerate perturbed eigenvalue problem
in which exact OSV rotations can be implemented  between the retained and discarded OSV subspaces. 
The simplicity of the OSV construction  leads to the block-diagonal structure of pair-specific OSV relaxation matrix
which decouples OSV rotations within a single orbital pair. 
The solution of pair-specific OSV relaxation elements enters the source of a single  Z-vector equation along with
the MO relaxation and the localization constraint,
as solved in a conventional way that is independent of the degrees of freedom.


The accuracy of this approach has been benchmarked on a set of  well studied molecules 
for optimized geometries and molecular dynamics simulations.
The OSV relaxation effects are significant and can be recovered
with the normal OSV selection for practical use of reproducing canonical RI-MP2
molecular structures.
Moreover, the classical molecular dynamics with OSV-MP2 input gradients
has been implemented. It has been demonstrated that using a normal OSV selection, 
all major peaks of the O-O/O-H radial distribution functions and vibrational densities of states
for protonated water tetramer and hexamer can be well identified.
A 200 ps well-tempered metadynamics simulation with OSV-MP2 gradients at 300 K 
has been shown to be capable of distinguishing the gauche and trans conformers of ethanol molecule.

There is much to explore for improving the current implementation by noting the
aspects as follows.  (1) Solving the Z-vector equation and two-electron integral
transformation therein in MO basis become one bottleneck step for large
molecules.  (2) The evaluation of energy gradients through Eq.~(\ref{eq:ec1e2e})
does not yet take advantage of OSV savings and therefore scales quickly with
system sizes.  (3) Embarrassing parallelization schemes seem obvious within the
OSV-MP2 framework by distributing local orbitals over many processes. (4)
Finally, the OSV-MP2 gradient computation is currently much slower than OSV-MP2
energy by 3-4 folds.  An appropriate scheme for pruning out insignificant OSV
relaxations and associated pairs shall further speed up gradient computation.
The efforts along these directions are being made and will be reported in
future.  

\begin{acknowledgement}

J.Y. acknowledges financial supports 
from the Hong Kong Research Grant Council (RGC)
Early Career Scheme (ECS) through Grant No. ECS27307517, 
and the Hui's fund provided by department of chemistry at the University of Hong Kong.
R.Y.Z. thanks Prof. Roberto Car for hosting his summer research and valuable discussions.
J.Y. thanks Dr. Qiming Sun for general assistance in PySCF package.

\end{acknowledgement}

\begin{suppinfo}

The file Supporting supporting.pdf contains further  results 
of the computations and is available free of charge.

\end{suppinfo}


\bibliography{manuscript}

\end{document}